\documentclass[12pt,a4paper]{article}
\usepackage{amssymb,amsmath}
\usepackage[dvips]{lscape,graphicx}

\newcommand{\bc}{\begin{center}}
\newcommand{\ec}{\end{center}}
\newcommand{\bd}{\begin{displaymath}}
\newcommand{\ed}{\end{displaymath}}
\newcommand{\be}{\begin{equation}}
\newcommand{\ee}{\end{equation}}
\newcommand{\ba}{\begin{array}}
\newcommand{\ea}{\end{array}}
\newcommand{\bt}{\begin{tabular}}
\newcommand{\et}{\end{tabular}}

\newcommand{\ds}{\displaystyle}

\begin{document}

\title{$E_6$ inspired composite Higgs model}

\author{R.~Nevzorov\footnote{On leave of absence from the Theory Department, SSC RF ITEP of NRC "Kurchatov Institute", Moscow, Russia.},
A.~W.~Thomas\\[5mm]
\itshape{ARC Centre of Excellence for Particle Physics at the Terascale and CSSM,}\\[0mm]
\itshape{Department of Physics, The University of Adelaide, Adelaide SA 5005, Australia}}

\date{}

\maketitle

\begin{abstract}{
\noindent
We consider a composite Higgs model embedded into a Grand Unified Theory
(GUT) based on the $E_6$ gauge group. The phenomenological viability of this $E_6$ inspired
composite Higgs model (E$_6$CHM) implies that standard model (SM) elementary fermions with
different baryon or lepton number should stem from different $27$ representations of $E_6$.
We present a six--dimensional orbifold GUT model in which the $E_6$ gauge symmetry is broken to the SM gauge
group so that the appropriate splitting of the bulk $27$--plets takes place. In this model the strongly
coupled sector is localised on one of the branes and possesses an $SU(6)$ global symmetry that
contains the $SU(3)_C\times SU(2)_W\times U(1)_Y$ subgroup. In this case the approximate
gauge coupling unification can be attained if the right--handed top quark is a composite state and
the elementary sector involves extra exotic matter beyond the SM which ensures anomaly cancellation.
The breakdown of the approximate $SU(6)$ symmetry at low energies in this model results in
a set of the pseudo--Nambu--Goldstone states which include a Higgs doublet and scalar colour
triplet. We discuss the generation of the masses of the SM fermions in the E$_6$CHM.
The presence of the TeV scale vector--like exotic quarks and scalar colour triplet may provide
spectacular new physics signals that can be observed at the LHC.
}
\end{abstract}

\newpage
\section{Introduction}
The properties of a new scalar particle, discovered by the ATLAS \cite{:2012gk} and
CMS \cite{:2012gu} collaborations at CERN, strongly suggest that it is the Higgs boson,
the particle related to the mechanism of the electroweak (EW) symmetry breaking (EWSB)
in the standard model (SM). Current data does not allow one to distinguish whether this
new scalar state is an elementary particle (up to very high energies) or a composite
state composed of more fundamental degrees of freedom. 
The idea of a composite Higgs boson, which was proposed in the 70's \cite{Terazawa:1976xx} 
and 80's \cite{composite-higgs}, implies that there exists a new,
strongly coupled sector. This sector generates the EW scale dynamically, in analogy with
the origin of the QCD scale. In such models the composite Higgs state has generically
a large quartic coupling and tends to be quite heavy. On the other hand, the recently observed
Higgs boson is sufficiently light, with mass around $m_h\simeq 125-126\,\mbox{GeV}$, that it corresponds
to a rather small value of the Higgs quartic coupling, $\lambda\simeq 0.13$. The relatively
low values of $m_h$ and $\lambda$ indicate that the Higgs field can emerge as a
pseudo--Nambu--Goldstone boson (pNGB) from the spontaneous breaking of an approximate
global symmetry of the strongly coupled sector. This idea was used before in the little--Higgs
models \cite{little-higgs}.

The pNGB Higgs idea is also realised in Randall--Sundrum (RS)
extra--dimensional scenarios, with the SM fields in the bulk \cite{Contino:2003ve}--\cite{Agashe:2004rs}.
Via the AdS/CFT correspondence, these scenarios are dual to the $4D$ composite Higgs
scenarios in which Kaluza--Klein excitations are associated with the low--lying bound
states at the compositeness scale, $f$ \cite{Contino:2003ve}--\cite{Contino:2006nn}.
Thus these models contain a sector of weakly--coupled elementary particles, including
the SM gauge bosons and SM fermions, as well as a second strongly interacting sector
resulting in a set of composite bound states that involves a Higgs doublet, massive excitations
of the elementary fields and so on. The elementary states couple weakly to the composite operators
of the strong sector. Because of this, at low energies
those states identified with SM fermions (bosons) are a mixture of the
corresponding elementary fermionic (bosonic) states and their vectorlike fermionic (bosonic)
composite partners. In this framework, which is known as partial compositeness \cite{Contino:2006nn, Kaplan:1991dc},
the SM states couple to the composite Higgs with a strength which is determined by the
fraction of the compositeness of this state. That is, for the effective up-- and down--quark
Yukawa couplings ($y^u_{ij}$ and $y^d_{ij}$ respectively) one gets
\begin{equation}
y^u_{ij} =s_q^i Y^{u}_{ij} s_u^j\,,\qquad y^d_{ij} =s_q^i Y^{d}_{ij} s_d^j\,,
\label{ch1}
\end{equation}
where $i,\,j=1,2,3$ run over three generations, $Y^{u}_{ij}$ and $Y^{d}_{ij}$ are the effective Yukawa
couplings of the composite Higgs field to the composite partners of the up-- and down--quarks,
while $s_u^j$ and $s_d^j$ are the fractions of compositeness of the right--handed SM quarks
of up-- and down--type, respectively, and $s_q^i$ are the fractions of compositeness of
the left--handed SM quarks. The couplings of the elementary states to the strongly interacting
sector explicitly break the global symmetry of the latter. As a consequence, the pNGB Higgs
potential arises from loops containing elementary states and that in turn leads to the suppression of the
effective quartic Higgs coupling.

The observed mass hierarchy in the quark and lepton sectors can be accommodated through
partial compositeness if the fractions of compositeness of the first and second generation
fermions are quite small; that is, the couplings of the corresponding elementary
states to their composite partners are very weak. Such weak couplings also substantially suppress
the flavor--changing effects and the modifications of the $W$ and $Z$ couplings associated with
the light SM fermions \cite{Contino:2006nn, Agashe:2004cp}, serving as a generalization of Glashow--Iliopoulos--Maiani
(GIM) mechanism of the SM \cite{Glashow:1970gm}. At the same time, the top quark is so heavy that
right--handed and left--handed top quarks ($t^c$ and $t$) should have sizeable fractions of
compositeness. Since precision data, such as $Z\to b\bar{b}$ measurements, imply that the
left--handed $b$--quark, and hence $t$, should have a reasonably small admixture of composite
partners, $t^c$ is expected to be almost completely composite.

If $t^c$ is entirely composite then the approximate unification of the SM gauge couplings, $\alpha_i$, can be achieved
very naturally \cite{Agashe:2005vg}. This happens, for example, when all composite objects fill in complete
$SU(5)$ representations and the sector of weakly--coupled elementary states involves
\begin{equation}
(q_i,\,d^c_i,\,\ell_i,\,e^c_i) + u^c_{\alpha} + \bar{q}+\bar{d^c}+\bar{\ell}+\bar{e^c}+\eta\,,
\label{ch2}
\end{equation}
where $\alpha=1,2$ runs over the first two generations and $i=1,2,3$ runs over all three.
We have denoted here the left-handed quark and lepton doublets by $q_i$ and $\ell_i$,
the right-handed up- and down-type quarks and charged leptons by $u_{\alpha}^c, d_i^c$ and
$e_i^c$, while the extra exotic states in Eq.~(\ref{ch2}), $\bar{q},\,\bar{d^c},\,\bar{\ell}$ and $\bar{e^c}$,
have exactly opposite $SU(3)_C\times SU(2)_W\times U(1)_Y$ quantum numbers to
left-handed quark doublets, right-handed down-type quarks, left-handed lepton doublets and
right-handed charged leptons, respectively. An extra exotic elementary state $\eta$ with spin $1/2$,
that does not participate in the $SU(3)_C\times SU(2)_W\times U(1)_Y$ gauge interactions,
is included to ensure the phenomenological viability of such a scenario. This scenario also implies
that the dynamics of the strongly interacting sector leads to the composite ${\bf 10} + {\bf \overline{5}} + {\bf 1}$
multiplets of $SU(5)$ which, in turn get combined with $\bar{q},\,\bar{d^c},\,\bar{\ell}$, $\bar{e^c}$ and $\eta$
forming a set of vector--like states. The only exceptions are the components of the $10$--plet
associated with the composite $t^c$, which survive down to the EW scale.

Using the one--loop renormalisation group equations (RGEs) it is rather easy to find the value
of $\alpha_3(M_Z)$ for which exact gauge coupling unification takes place in this model
\be
\ds\frac{1}{\alpha_3(M_Z)}=\frac{1}{b_1-b_2}\biggl[\ds\frac{b_1-b_3}{\alpha_2(M_Z)}-
\ds\frac{b_2-b_3}{\alpha_1(M_Z)}\biggr]\,,
\label{ch3}
\ee
where $b_i$ are one--loop beta functions, with the indices $1,\,2,\,3$ corresponding
to the $U(1)_Y$, $SU(2)_W$ and $SU(3)_C$ interactions. Since all composite states come in complete
$SU(5)$ multiplets, the strong sector does not contribute to the differential running, which is determined
by $(b_i-b_j)$ in the one--loop approximation.
Then, for $\alpha(M_Z)=1/127.9$, $\sin^2\theta_W=0.231$ and the elementary particle spectrum given by
Eq.~(\ref{ch2}), one finds that exact gauge coupling unification can be obtained for
$\alpha_3(M_Z)\simeq 0.109$\,. Although this value of $\alpha_3(M_Z)$ is substantially lower than
the central measured low energy value, this result does indicate that at high energies
an approximate gauge coupling unification can be attained within the composite Higgs model
with a composite $t^c$. It was argued that the inclusion of higher order effects coming from
the strongly coupled sector, as well as the weakly--coupled elementary sector, may improve the gauge
coupling unification, which takes place around the scale $M_X\sim 10^{15}-10^{16}\, \mbox{GeV}$
in these models \cite{Agashe:2005vg}--\cite{Barnard:2014tla}.

In this context it is especially interesting to consider the embedding of the composite Higgs models
into well known Grand Unified Theories (GUTs) in which all elementary quark and lepton states are
the components of some irreducible representation of the GUT gauge group. Here we focus on the
$E_6$ gauge theory. In this GUT all elementary SM fermions can originate from the fundamental
$27$-dimensional representation of $E_6$. To suppress baryon and lepton number violating operators,
that lead to rapid proton decay and too large masses of the left--handed neutrinos, the low energy
effective Lagrangian of this $E_6$ inspired composite Higgs model (E$_6$CHM) has to be invariant
with respect to the global $U(1)_B$ and $U(1)_L$ symmetries associated with the conservation of
baryon and lepton numbers, to a very good approximation. In the simplest case this implies that elementary
quark and lepton fields with different baryon or lepton number should come from different $27$--plets,
whereas all other components of these $27$--plets acquire masses somewhat close to the scale $M_X$
where the $E_6$ gauge symmetry is broken down to the SM gauge group. The corresponding splitting
of the $27$--plets can take place in the orbifold GUTs.

The layout of this paper is as follows. In Section 2 we present a six--dimensional ($6D$) orbifold GUT model
based on the $E_6$ gauge group in which $E_6$ is broken down to $SU(3)_C\times SU(2)_W\times U(1)_Y$
gauge symmetry, so that all SM fermions with different baryon or lepton number stem from different
fundamental representations of $E_6$. At low energies the weakly--coupled elementary sector of this model
involves a set of states given by Eq.~(\ref{ch2}). In this SUSY GUT model all fields of the strongly interacting
sector reside on the brane where $E_6$ gauge symmetry is broken down to $SU(6)$. This $SU(6)$ symmetry
includes $SU(3)_C\times SU(2)_W\times U(1)_Y$ subgroup. Here we assume that $SU(6)$ remains an
approximate global symmetry of the strongly coupled sector even at low energies. This can happen when
the gauge couplings of the strongly interacting sector are considerably larger than the SM gauge couplings
below the scale $M_X$. Assuming that the breakdown of the approximate global $SU(6)$ symmetry down to its
$SU(5)$ subgroup takes place at low energies, the spectrum of the E$_6$CHM involves a set of the
pseudo--Nambu--Goldstone states, including a composite Higgs doublet and scalar colour triplet. In section 3
we discuss the generation of masses of the SM fermions and other phenomenological implications of the E$_6$CHM.
Our results are summarized in Section 4.

\section{$E_6$ orbifold GUT model in six dimensions}

Higher--dimensional theories offer new possibilities for gauge symmetry breaking. A simple and elegant
scheme is provided by orbifold compactifications which have been considered for SUSY GUT models in five
dimensions \cite{5d-susy-ogut}--\cite{Braam:2010sy} and six dimensions \cite{5d+6d-susy-ogut}--\cite{Buchmuller:2004eg}.
These models apply ideas that first appeared in string--motivated work \cite{Candelas:1985en}, where it was pointed out that
the gauge symmetry could be broken by identifications imposed on the gauge fields under the spacetime symmetries of an orbifold.
More recently, orbifold compactifications of the heterotic string have been constructed which can account for the SM in four dimensions
and which have five--dimensional or six--dimensional GUT structures as intermediate steps, very similar to orbifold GUT
models \cite{Buchmuller:2005jr}. In the context of Sherk-Schwarz compactification the models of
composite quarks and leptons were discussed in \cite{Chaichian:2001fs}.

In this section we study an $N=1$ supersymmetric (SUSY) GUT in $6D$ that can lead at low energies
to the field content of the weakly--coupled elementary sector given by Eq.~(\ref{ch2}). In particular,
we focus on the SUSY GUT based on the $E_6\times G_0$ gauge group. This SUSY GUT implies that at
high energies $E_6$ and $G_0$ are broken down to their subgroups, i.e. $SU(3)_C\times SU(2)_W\times U(1)_Y$
and $G$, respectively, which are associated with the elementary and strongly coupled sectors.
Fields from the strongly coupled sector can be charged under both the $E_6$ and $G_0$ gauge groups,
while the elementary states participate in the $E_6$ interactions only.

We further assume that all elementary quark and lepton fields are components of the bulk
$27$ supermultiplets of $E_6$. In the four--dimensional $N=1$ SUSY models based on the $E_6$ gauge
group, the fundamental 27-dimensional representation involves components $\Phi_i$ that correspond to the
left-handed quark and lepton supermultiplets ($q_i$ and $\ell_i$), right--handed up- and down-type quark
supermultiplets ($u_i^c$ and $d_i^c$), right--handed charged and neutral lepton superfields ($e^c_i$ and
$\nu^c_i$), a SM singlet superfield $s_i$, charged $\pm 1/3$ exotic quark supermultiplets ($h^c_i$ and $h_i$)
as well as two $SU(2)_W$ doublet superfields ($h^{u}_{i}$ and $h^{d}_{i}$) that do not carry baryon or
lepton number. The minimal $N=1$ supersymmetry in $6D$ corresponds to $N=2$ in $4D$. Indeed, because
a $6D$ fermion state is composed of two $4D$ Weyl fermions, $\psi_i$ and $\psi^c_i$, SUSY implies that each
$6D$ superfield includes two complex scalars, $\phi_i$ and $\phi^c_i$, as well. The fields $\psi_i, \psi^c_i, \phi_i$
and $\phi^c_i$ form a $4D$ $N=2$ hypermultiplet which involves two $4D$ $N=1$ chiral superfields:
$\Phi_i=(\phi_i,\,\psi_i)$ and its conjugate $\overline{\Phi}_i = (\phi^c_i,\,\psi^c_i)$, with opposite quantum
numbers. Thus each bulk $27$ supermultiplet $\widehat{\Phi}_i$ contains two $4D$ $N=1$
supermultiplets, $27$ and $\overline{27}$.

The $E_6$ gauge supermultiplet that exists in the bulk should contain vector bosons
$A_{M}$ ($M=0,1,2,3,5,6$) and $6D$ Weyl fermions (gauginos). The $6D$ gauginos are composed
of two $4D$ Weyl fermions, $\lambda$ and $\lambda'$. These fields can be grouped into vector
and chiral multiplets of the $N=1$ supersymmetry in $4D$, i.e.
\be
V=(A_{\mu}, \lambda)\,,\qquad\qquad \Sigma=\biggl((A_5+i A_6)/\sqrt{2},\lambda'\biggr)\,,
\label{ch4}
\ee
where $V$, $A_{M}$, $\lambda$ and $\lambda'$ are matrices in the adjoint
representation of $E_6$ and $\mu=0,1,2,3$. These two $N=1$ supermultiplets also form an $N=2$
vector supermultiplet in $4D$.

We consider the compactification of two extra dimensions on a torus $T^2$ with two fixed radii,
$R_5$ and $R_6$. Thus two extra dimensions $y (=x_5)$ and $z (=x_6)$ are compact, i.e.
$y\in (-\pi R_5, \pi R_5]$ and $z\in (-\pi R_6, \pi R_6]$. The sizes of the radii, $R_5$ and $R_6$,
are determined by the GUT scale, $M_X$. The orbifold $T^2/Z_2$ is obtained by dividing the torus
$T^2$ with a $Z_2$ transformation which acts on $T^2$ according to $y\to -y$ and $z\to -z$.
The components of the bulk supermultiplets transform under the $Z_2$ action as well.
The Lagrangian is invariant under the $Z_2$ transformation. The orbifold $T^2/Z_2$ has the
following set of fixpoints: $(0,0)$, $(\pi R_5,0)$, $(0,\pi R_6)$ and $(\pi R_5,\pi R_6)$.
The $Z_2$ transformation can be regarded as an equivalence relation that allows one to reduce
the physical region associated with the compactification on the orbifold $T^2/Z_2$ to a pillow
with the four fixed points of the $Z_2$ transformation as corners.

\subsection{The breakdown of $E_6$ to $SU(4)'\times SU(2)_W \times SU(2)_N \times U(1)'$}

Here we examine $6D$ SUSY GUT compactified on the orbifold $T^2/(Z_2 \times Z^{I}_2 \times Z^{II}_2)$.
The $Z_2$, $Z^{I}_2$ and $Z^{II}_2$ symmetries are reflections. The $Z_2$ symmetry transformation is
defined as before, i.e. $y\to -y$, $z\to -z$. The transformation associated with the reflection $Z^{I}_2$ is
given by $y'\to -y'$, $z\to -z$, where $y' = y - \pi R_5/2$. The reflection $Z^{II}_2$ acts as $y\to -y$, $z'\to -z'$,
with $z' = z - \pi R_6/2$. The additional $Z^{I}_2$ and $Z^{II}_2$ reflection symmetries introduce extra fixed
points that lead to the further reduction of the physical region which is limited by these fixed points.
In this case the physically irreducible space is a pillow, in which $y\in [0, \pi R_5/2]$ and $z\in [0, \pi R_6/2]$,
with the four $4D$ walls (branes) located at its corners.

The consistency of the construction requires that the Lagrangian of the orbifold SUSY GUT model under
consideration is invariant under $Z_2$, $Z^{I}_2$ and $Z^{II}_2$ reflections. Each reflection symmetry,
$Z_2$, $Z^{I}_2$ and $Z^{II}_2$, has its own orbifold parity, $P$, $P_{I}$ and $P_{II}$.
To ensure the invariance of the Lagrangian, the components $\Phi_i$ and $\overline{\Phi}_i$ of the bulk
$27$ supermultiplet should transform under $Z_2$, $Z^{I}_2$ and $Z^{II}_2$ as follows
\be
\ba{ll}
\Phi_i (x, -y, -z) = P_{ii} \Phi_i(x, y, z)\,,&\qquad  \overline{\Phi}_i(x, -y, -z) = -P_{ii} \overline{\Phi}_{i}(x, y, z)\,,\\
\Phi_i(x, -y', -z) = P^{I}_{ii} \hat{\Phi}_i(x, y', z)\,,&\qquad  \overline{\Phi}_i(x, -y', -z) = -P^{I}_{ii} \overline{\Phi}_i(x, y', z)\,,\\
\Phi_i(x, -y, -z') = P^{II}_{ii} \hat{\Phi}_i(x, y, z')\,,&\qquad  \overline{\Phi}_i(x, -y, -z') = -P^{II}_{ii} \overline{\Phi}_i(x, y, z')\,,
\ea
\label{ch5}
\ee
where $P$, $P_{I}$ and $P_{II}$ are diagonal matrices with eigenvalues $\pm 1$ that act on each component
of the fundamental representation of $E_6$, making some components positive and some components negative.

One can specify the matrix representation of the orbifold parity assignments in terms of the $E_6$
weights $\alpha_{i}$ and gauge shifts, $\Delta$, $\Delta^{I}$ and $\Delta^{II}$ corresponding to $Z_2$,
$Z^{I}_2$ and $Z^{II}_2$. Then the diagonal elements of the matrices $P$, $P^{I}$ and $P^{II}$
can be written in the following form \cite{Braam:2010sy}
\be
\ba{c}
(P)_{ii}=\sigma\exp\{2\pi i \Delta \alpha_i\}\,,\qquad\qquad
(P^{I})_{ii}=\sigma_{I}\exp\{2\pi i \Delta^{I} \alpha_i\}\,,\\
(P^{II})_{ii}=\sigma_{II}\exp\{2\pi i \Delta^{II} \alpha_i\}\,,
\ea
\label{ch6}
\ee
where $\sigma$, $\sigma_{I}$ and $\sigma_{II}$ are parities of the
bulk $27$ supermultiplet, i.e. $\sigma, \sigma_{I}, \sigma_{II} \in \{+,-\}$.
In the case of the fundamental representation of $E_6$ the particle assignments
of the weights are well known (see, for example \cite{Braam:2010sy}). Here we choose
the following gauge shifts
\be
\ba{c}
\Delta=\biggl(0,\,0,\,0,\,\dfrac{1}{2},\,0,\,0\biggr)\,,\qquad
\Delta^{I}=\biggl(\dfrac{1}{2},\,\dfrac{1}{2},\,\dfrac{1}{2},\,\dfrac{1}{2},\,\dfrac{1}{2},\,0\biggr)\,,\\
\Delta^{II}=\biggl(\dfrac{1}{2},\,\dfrac{1}{2},\,\dfrac{1}{2},\,0,\,\dfrac{1}{2},\,0\biggr)\,,
\ea
\label{ch7}
\ee
that correspond to the orbifold parity assignments shown in Table~\ref{tab1}.

\begin{table}[ht]
\centering
\begin{tabular}{|c|c|c|c|c|c|c|c|c|c|c|c|}
\hline
                  & $q$ & $d^c$ & $u^c$ & $\ell$ & $e^c$ & $\nu^c$ & $h^u$ & $h^d$ & $h$ & $h^c$ & $s$\\
\hline
$Z_2$        & $+$&  $-$      & $+$    &  $-$   &  $+$    &    $-$      &  $+$     &    $-$   & $+$ &  $-$    & $-$ \\
\hline
$Z_2^{I}$ & $-$ &   $+$    & $+$    & $-$    & $+$     &    $+$    &  $-$       &    $-$   & $+$ &  $+$   & $+$ \\
\hline
$Z_2^{II}$& $-$ &    $-$     & $+$    & $+$   &  $+$   &     $-$     &   $-$      &   $+$   & $+$ &   $-$   & $-$ \\
\hline
$Z_2^{III}$& $+$ & $+$    &  $+$   & $+$   & $+$    &     $+$    &  $+$      &   $+$  & $+$  &  $+$  & $+$ \\
\hline
\end{tabular}
\caption{Orbifold parity assignments in the bulk $27$ supermultiplet
with $\sigma=\sigma_{I}=\sigma_{II}=\sigma_{III}=+1$.}
\label{tab1}
\end{table}

The supermultiplets $V$ and $\Sigma$, which are components of the $E_6$ gauge supermultiplet,
transform under $Z_2$, $Z^{I}_2$ and $Z^{II}_2$ as follows
\be
\ba{ll}
V(x, -y, -z) = P V(x, y, z) P^{-1},&
\Sigma(x, -y, -z) = -P \Sigma (x, y, z) P^{-1},\\
V(x, -y', -z) = P^{I} V(x, y', z) (P^{I})^{-1},&
\Sigma(x, -y', -z) = -P^{I} \Sigma (x, y', z) (P^{I})^{-1},\\
V(x, -y, -z') = P^{II} V(x, y, z') (P^{II})^{-1},&
\Sigma(x, -y, -z') = -P^{II} \Sigma (x, y, z') (P^{II})^{-1}.
\ea
\label{ch8}
\ee
In Eq.~(\ref{ch8}) $V(x, y, z)=V^{A}(x, y, z) T^{A}$ and $\Sigma(x, y, z)=\Sigma^A(x, y, z) T^A$
where $T^A$ is the set of generators of the $E_6$ group. Since different components of the
bulk supermultiplets transform differently under $Z_2$, $Z^{I}_2$ and $Z^{II}_2$ reflections,
the $4D$ $N=2$ supersymmetry is broken down to $4D$ $N=1$ SUSY. Moreover, $E_6$ gauge
symmetry is also broken by the parity assignments specified in Table~\ref{tab1}, because
$P$, $P^{I}$ and $P^{II}$ are not unit matrices and do not commute with all $E_6$ generators.

On the brane $O$, situated near the the fixed point $y=z=0$ which is associated with the $Z_2$ reflection
symmetry, the $E_6$ gauge symmetry is broken down to $SU(6)\times SU(2)_N$. This follows from
the $P$ parity assignment in the bulk $27$ supermultiplet. The $27$--plet of $E_6$ decomposes under
the $SU(6)\times SU(2)_N$ subgroup as follows:
$$
27\to (15,\, 1) + (\overline{6},\,2)\,,
$$
where the first and second quantities in brackets are the $SU(6)$ and $SU(2)_N$ representations.
The multiplet $(\overline{6},\,2)$ involves two $SU(3)_C$ triplets $d^c$ and $h^c$, two $SU(2)_W$
doublets $\ell$ and $h^d$ as well as two SM singlets $\nu^c$ and $s$, which are contained in the $27$--plet.
In this case the SM gauge group is a subgroup of $SU(6)$, whereas none of the $SU(2)_N$ gauge bosons
participate in the $SU(3)_C\times SU(2)_W\times U(1)_Y$ gauge interactions. From Table~\ref{tab1}
one can see that the components of the $27$ supermultiplet, that correspond to the multiplet
$(\overline{6},\,2)$, transform differently under the $Z_2$ symmetry as compared with the other
components of the $27$--plet which form the $(15,\, 1)$ representation of $SU(6)$. We further assume
that all fields from the strongly coupled sector reside on the $O$ brane.

The $P^{I}$ parity assignment associated with the $Z_2^{I}$ symmetry leads to the breakdown of
the $E_6$ gauge group to the $SU(6)'\times SU(2)_W$ subgroup on the brane $O_{I}$ located
at the fixed point $y=\pi R_5/2$, $z=0$. Indeed, according to Table~\ref{tab1} all $SU(2)_W$
doublet components of the bulk $27$ supermultiplet, which form the $(6,\,2)$ representation, transform
differently under the $Z^{I}_2$ reflection, as compared with all other components of this
supermultiplet which compose $(\overline{15},\, 1)$ of $SU(6)'$. In this case the $SU(3)_C$ symmetry
is a subgroup of $SU(6)'$. To ensure the breakdown of the $E_6$ gauge symmetry to the SM gauge
group, we assume that two pairs of the supermultiplets $(15,\, 1)$ and $(\overline{15},\, 1)$ of $SU(6)'$
are confined on the brane $O_{I}$.

The $E_6$ symmetry is also broken on the brane $O_{II}$ placed at the fixed point $y=0$, $z=\pi R_6/2$
of the $Z_2^{II}$ reflection symmetry. The $P_{II}$ parity assignment is such that the 16 components
$q$, $d^c$, $\nu^c$, $h^u$, $h^c$ and $s$ of the bulk $27$--plet are odd, while all other components
are even. Because the symmetry breaking mechanism in the orbifold GUT models preserves the rank
of the group, the unbroken subgroup at the fixed point $O_{II}$ should be $SO(10)'\times U(1)'$.
Indeed, the $16$ components of the bulk $27$--plet mentioned above constitute a $16$--dimensional spinor
representation of $SO(10)'$. The other $10$ components $u^c$, $\ell$, $h^d$ and $h$ of  the bulk $27$--plet
form a $10$--dimensional vector representation of $SO(10)'$, whereas the $e^c$ component represents
an $SO(10)'$ singlet. The $SU(3)_C$ and $SU(2)_W$ groups are subgroups of $SO(10)'$. It is worth noting that
ordinary $SO(10)$ and $SO(10)'$ are not the same subgroups of $E_6$. In particular, the 16-plets of
$SO(10)$ and $SO(10)'$ are formed by different components of the fundamental representation of $E_6$.
The $U(1)'$ charges of the different components of the $27$--plet are given in Table~\ref{tab2}.
The consistency of the orbifold GUT model under consideration requires that three pairs of $e^c_i$ and
$\overline{e^c_i}$ superfields as well as $45$--dimensional representations of $SO(10)'$ reside on the
brane $O_{II}$.

\begin{table}[ht]
  \centering
  \begin{tabular}{|c|c|c|c|c|c|c|c|c|c|c|c|}
    \hline
    & $q$ & $d^c$ & $u^c$ & $\ell$ & $e^c$ & $\nu^c$ & $h^u$ & $h^d$ & $h$ & $h^c$ & $s$\\
 \hline
$\sqrt{24}Q'_i$ & $1$ & $1$ & $-2$ & $-2$ & $4$ & $1$ & $1$ & $-2$ & $-2$ & $1$ & $1$\\
 \hline
$\sqrt{24}\widetilde{Q}_i$ & $1$ & $-1$ & $2$ & $0$ & $0$ & $3$ & $-3$ & $0$ & $-2$ & $-1$ & $3$ \\
 \hline
$\sqrt{{40}}Q^{N}_i$ & $1$ & $2$ & $1$ & $2$ & $1$ & $0$ & $-2$ & $-3$ & $-2$ & $-3$ & $5$ \\
\hline
$\sqrt{\frac{5}{3}}Q^{Y}_i$ & $\frac{1}{6}$ & $\frac{1}{3}$ &$-\frac{2}{3}$ &  $-\frac{1}{2}$
& $1$ & $0$ & $\frac{1}{2}$ & $-\frac{1}{2}$ & $-\frac{1}{3}$ &
 $\frac{1}{3}$ & $0$ \\
 \hline
  \end{tabular}
  \caption{The $U(1)'$, $\widetilde{U}(1)$, $U(1)_N$ and $U(1)_Y$ charges ($Q'_i$, $\widetilde{Q}_i$, $Q^{N}_i$ and $Q^{Y}_i$ respectively)
  of the different components of the $27$--plet.}
  \label{tab2}
\end{table}

In addition to the three branes mentioned above, there is a fourth brane $O_{III}$ which is situated at the corner
$y=\pi R_5/2$, $z=\pi R_6/2$ of the physically irreducible space. The $Z^{III}_2$ reflection corresponding to
this brane is obtained by combining the three symmetries $Z_2$, $Z^{I}_2$ and $Z^{II}_2$. The corresponding parity
assignment $P_{III}=P\, P_{I}\, P_{II}$. Combining three parity assignments $P$, $P^{I}$ and $P^{II}$ one can see
that $P^{III}$ is just an identity matrix. This implies that on the brane $O_{III}$ the $E_6$ gauge symmetry remains
intact, while $N=2$ supersymmetry is broken to $N=1$ SUSY. The consistency of this orbifold GUT model requires
that two $27$--plets are confined on this brane.

The unbroken gauge group of the low-energy effective $4D$ theory is given by the intersection of the $E_6$ subgroups
at the fixed points $O$, $O_{I}$, $O_{II}$ and $O_{III}$. The intersection of $SU(6)\times SU(2)_N$, $SU(6)'\times SU(2)_W$
and $SO(10)'\times U(1)'$ yields the group $SU(4)'\times SU(2)_W \times SU(2)_N \times U(1)'$. The $SU(3)_C$ group is a subgroup
of $SU(4)'$, which is in turn a subgroup of $SO(10)'$.

\subsection{The breakdown of $SU(4)'\times SU(2)_W \times SU(2)_N \times U(1)'$ to the SM gauge group}

As it follows from the Table~\ref{tab1}, in general the bulk $27$--plets include components with even and odd parities,
$P$, $P^{I}$, $P^{II}$ and $P^{III}$. However, only components for which all parities are positive are allowed to
have zero--modes. This means that the corresponding fields can survive below the GUT scale $M_X$. None of the other components
of the bulk $27$--plets possess massless modes. The elementary states, $u^c_{\alpha}$, $e^c_i$ and $\bar{e^c}$,
can originate from the bulk $27$ supermultiplets $\widehat{\Phi}^{u}_{i}$,  $\widehat{\Phi}^{\overline{u}}_{i}$,
$\widehat{\Phi}^{e}_{i}$ and  $\widehat{\Phi}^{\overline{e}}_{i}$ that decompose as follows
\be
\ba{c}
\widehat{\Phi}^{u}_{i} = \ds\left(27,\,+,\,+,\,+,\,+\right),\qquad
\widehat{\Phi}^{\overline{u}}_{i} = \ds\left(27,\,-,\,-,\,-,\,-\right),\\
\widehat{\Phi}^{e}_{i} = \ds\left(27,\,+,\,+,\,+,\,+\right),\qquad
\widehat{\Phi}^{\overline{e}}_{i} = \ds\left(27,\,-,\,-,\,-,\,-\right),
\ea
\label{ch9}
\ee
where the quantities in brackets are the $E_6$ representation as well as the values of $\sigma$, $\sigma_{I}$, $\sigma_{II}$ and
$\sigma_{III}$ associated with this representation. In Eq.~(\ref{ch9}) $i=1,2,3$ as before. The parities of $\widehat{\Phi}^{u}_{i}$
are chosen so that their $u^c_i$, $e^c_i$ and $h_i$ components of the $N=1$ chiral supermultiplet, $\Phi^{u}_i$, have positive
parities with respect to all reflection symmetries (see Table~\ref{tab1}). Because the invariance of the $6D$ action requires
that the parities of the $4D$ chiral supermultiplets $\Phi^{u}_i$ and $\overline{\Phi}^{u}_i$ are opposite, the $N=1$ chiral
supermultiplet $\overline{\Phi}^{u}_i$ does not contain any even components. On the other hand, in the case of
$\widehat{\Phi}^{\overline{u}}_{i}$ only components $\overline{u^c}_i$, $\overline{e^c}_i$ and $\overline{h}_i$ of
the $N=1$ chiral supermultiplet $\overline{\Phi}^{\overline{u}}_i$ are even. Thus the Kaluza--Klein (KK) expansion of
the bulk supermultiplets $\widehat{\Phi}^{u}_{i}$ and $\widehat{\Phi}^{\overline{u}}_{i}$ contains zero modes that
form $N=1$ chiral supermultiplets with the quantum numbers of $u^c_i$, $e^c_i$, $h_i$, $\overline{u^c}_i$, $\overline{e^c}_i$
and $\overline{h}_i$. Similar zero modes come from the KK expansion of $\widehat{\Phi}^{e}_{i}$ and
$\widehat{\Phi}^{\overline{e}}_{i}$.

Here we assume that one component, $\varphi$, of the $45$--dimensional representations of $SO(10)'$ localised on the
brane $O_{II}$, which is associated with the Cartan algebra generator of the $\widetilde{U}(1)$ subgroup of $SO(10)'$
(see Table~\ref{tab2}), acquires a non-zero vacuum expectation value (VEV), $\varphi_0$, which is somewhat smaller
than the GUT scale, breaking the $SU(4)'\times SU(2)_W \times SU(2)_N \times U(1)'$ gauge group down to
$SU(3)_C \times SU(2)_W \times SU(2)_N \times \widetilde{U}(1) \times U(1)'$. If this superfield couples to
$\widehat{\Phi}^{e}_{i}$ and $\widehat{\Phi}^{\overline{e}}_{i}$ then the zero modes $u^c_i$ and $\overline{u^c}_i$
as well as $h_i$ and $\overline{h}_i$ gain large masses ($\sim \varphi_0$), forming vectorlike states. Therefore only $N=1$ chiral
superfields with the quantum numbers of $e^c_i$ and $\overline{e^c}_i$ remain massless in this case. To forbid
any couplings of $\varphi$ to other bulk supermultiplets one can impose a discrete $Z_2^u$ symmetry
under which $\varphi$ and $\widehat{\Phi}^{e}_{i}$ are odd whereas all other supermultiplets are even.
Since the superfield $\varphi$ resides on the brane $O_{II}$, it does not interact with the superfields
localised on the brane $O$. Therefore the $SU(6)\times SU(2)_N$ global symmetry of the strongly coupled sector
remains intact.

Another discrete $Z_2^e$ symmetry, under which only three superfields $e^c_i$, which are confined on brane $O_{II}$,
and $\widehat{\Phi}^{\overline{u}}_{i}$ are odd while all other supermultiplets are even, allows us to suppress the
interaction between $e^c_i$ and the corresponding components of all bulk supermultiplets except $\widehat{\Phi}^{\overline{u}}_{i}$.
In this case the superfield $e^c_i$ gets combined with the appropriate zero modes of $\widehat{\Phi}^{\overline{u}}_{i}$
so that the resulting vectorlike states gain masses of order of $M_X$. As a consequence, only zero modes associated with
the $\overline{u^c}_i$ and $\overline{h}_i$ components of $\widehat{\Phi}^{\overline{u}}_{i}$
remain massless. In addition, we impose a $Z_2^{\overline{e}}$ symmetry which implies that $\overline{e^c_i}$ and
$\widehat{\Phi}^{u}_{i}$ are odd while all other supermultiplets are even. This symmetry allows for the formation of
vectorlike states which are formed by the components of the superfield $\overline{e^c_i}$ and the appropriate
zero modes of $\widehat{\Phi}^{u}_{i}$. Again in general these states gain masses of order of $M_X$. Because of
this, the set of the zero modes involves only the $u^c_i$ and $h_i$ components of $\widehat{\Phi}^{u}_{i}$.

The $4D$ supermultiplets $q_i$, $d^c_i$, $\bar{q}$ and $\bar{d^c}$ can stem from another six pairs of the
bulk $27$--plets
\be
\ba{c}
\widehat{\Phi}^{q}_{i} = \ds\left(27,\,+,\,-,\,-,\,+\right),\qquad
\widehat{\Phi}^{\overline{q}}_{i} = \ds\left(27,\,-,\,+,\,+,\,-\right),\\
\widehat{\Phi}^{d}_{i} = \ds\left(27,\,-,\,+,\,-,\,+\right),\qquad
\widehat{\Phi}^{\overline{d}}_{i} = \ds\left(27,\,+,\,-,\,+,\,-\right).
\ea
\label{ch10}
\ee
Using the orbifold parity assignments presented in Table~\ref{tab1}, one can check that all parities of
$q_i$ and $h^u_i$ components of $\widehat{\Phi}^{q}_{i}$, $\overline{q_i}$ and $\overline{h^u_i}$ components
of $\widehat{\Phi}^{\overline{q}}_{i}$ as well as $d^c_i$, $\nu^c_i$, $h^c_i$ and $s_i$ components of
$\widehat{\Phi}^{d}_{i}$ and $\overline{d^c_i}$, $\overline{\nu^c_i}$, $\overline{h^c_i}$ and $\overline{s_i}$
components of $\widehat{\Phi}^{\overline{d}}_{i}$ are positive so that the KK expansions of the corresponding
$6D$ superfields should contain the appropriate zero modes. Finally, in order to get $4D$ supermultiplets $\ell_i$ and
$\bar{\ell}$ the set of the bulk $27$--plets should be supplemented by
\be
\ba{c}
\widehat{\Phi}^{\ell}_{i} = \ds\left(27,\,-,\,-,\,+,\,+\right),\qquad
\widehat{\Phi}^{\overline{\ell}}_{i} = \ds\left(27,\,+,\,+,\,-,\,-\right).
\ea
\label{ch11}
\ee
The set of zero modes of $\widehat{\Phi}^{\ell}_{i}$ and $\widehat{\Phi}^{\overline{\ell}}_{i}$ involves $N=1$ chiral
supemultiplets with the quantum numbers of $\ell_i$, $h^d_i$, $\overline{\ell_i}$ and $\overline{h^d_i}$. The complete
set of the bulk $27$--plets and their zero modes, which survive below $\langle \varphi \rangle = \varphi_0$, are specified
in Table~\ref{tab3}. It is assumed that the mass terms involving zero modes of the $6D$ supermultiplets with exactly opposite
$SU(3)_C\times SU(2)_W\times U(1)_Y$ quantum numbers are not allowed. Such mass terms can be forbidden
by the $Z^b_2$ symmetry, under which  $\widehat{\Phi}^{u}_{i}$, $\widehat{\Phi}^{e}_{i}$, $\widehat{\Phi}^{q}_{i}$,
$\widehat{\Phi}^{d}_{i}$, $\widehat{\Phi}^{\ell}_{i}$ are even, whereas $\widehat{\Phi}^{\overline{u}}_{i}$,
$\widehat{\Phi}^{\overline{e}}_{i}$, $\widehat{\Phi}^{\overline{q}}_{i}$, $\widehat{\Phi}^{\overline{d}}_{i}$,
$\widehat{\Phi}^{\overline{\ell}}_{i}$ are odd.

\begin{table}[ht]
\centering
\begin{tabular}{|c|c|c|c|c|c|c|c|c|c|c|}
\hline
 & $\widehat{\Phi}^{u}_{i}$ & $\widehat{\Phi}^{e}_{i}$ & $\widehat{\Phi}^{q}_{i}$ & $\widehat{\Phi}^{d}_{i}$ & $\widehat{\Phi}^{\ell}_{i}$ & $\widehat{\Phi}^{\overline{u}}_{i}$
 & $\widehat{\Phi}^{\overline{e}}_{i}$ & $\widehat{\Phi}^{\overline{q}}_{i}$ & $\widehat{\Phi}^{\overline{d}}_{i}$ & $\widehat{\Phi}^{\overline{\ell}}_{i}$ \\
\hline
$E \lesssim M_X$ & $u^c_i$, $e^c_i$, & $u^c_i$, $e^c_i$, & $q_i$, & $d^c_i$, $\nu^c_i$, & $\ell_i$,   &$\overline{u^c}_i$, $\overline{e^c}_i$,
& $\overline{u^c}_i$, $\overline{e^c}_i$, & $\overline{q_i}$, & $\overline{d^c_i}$, $\overline{\nu^c_i}$, & $\overline{\ell_i}$,  \\
& $h_i$ & $h_i$  & $h^u_i$  & $h^c_i$, $s_i$ & $h^d_i$ & $\overline{h}_i$  & $\overline{h}_i$ & $\overline{h^u_i}$  & $\overline{h^c_i}$, $\overline{s_i}$ & $\overline{h^d_i}$\\
\hline
$E \lesssim \varphi_0$ & $u^c_i$,  &  $e^c_i$, & $q_i$, & $d^c_i$, $\nu^c_i$, & $\ell_i$,   &$\overline{u^c}_i$,
& $\overline{e^c}_i$, & $\overline{q_i}$, & $\overline{d^c_i}$, $\overline{\nu^c_i}$, & $\overline{\ell_i}$,  \\
& $h_i$ &   & $h^u_i$  & $h^c_i$, $s_i$ & $h^d_i$ & $\overline{h}_i$  &  & $\overline{h^u_i}$  & $\overline{h^c_i}$, $\overline{s_i}$ & $\overline{h^d_i}$\\
\hline
$E \lesssim \phi_0$ & $u^c_i$  &  $e^c_i$ & $q_i$ & $d^c_i$ & $\ell_i$  &$\overline{u^c}$
& $\overline{e^c}$ & $\overline{q}$ & $\overline{d^c}$ & $\overline{\ell}$  \\
\hline
\end{tabular}
\caption{The components of the bulk $27$--plets that survive below the scales $M_X$, $\varphi_0$ and $\phi_0$.
The index $i=1,2,3$ runs over all three generations.}
\label{tab3}
\end{table}

The $6D$ supermultiplets mentioned above lead to the set of zero modes which compose three pairs of complete
$N=1$ chiral $27$ and $\overline{27}$--plets. Two $\overline{27}$ supermultiplets (say, associated with $i=1,2$)
can get combined with two $27$--plets which are located on the $O_{III}$ brane resulting in the set of vectorlike states.
The mass scale $M_0$ associated with the masses of these states can be chosen slightly lower than $\varphi_0$.
We also assume that the $\nu^c$ and $\overline{\nu^c}$
components of one pair of  $\overline{15}$ and $15$ of $SU(6)'$, as well as the $s$ and $\overline{s}$ components of
another pair of $\overline{15}$ and $15$ of $SU(6)'$ localised on the brane $O_{I}$, acquire non--zero VEVs of order
of $\phi_0$ but somewhat below $M_0$ and $\varphi_0$. The VEVs of  $\nu^c$ and $\overline{\nu^c}$ break
$SU(3)_C \times SU(2)_W \times SU(2)_N \times \widetilde{U}(1) \times U(1)'$ gauge symmetry down to
the $SU(3)_C \times SU(2)_W \times U(1)_Y \times U(1)_N$ subgroup\footnote{Different phenomenological aspects of
SUSY models with extra $U(1)_N$ gauge symmetry were considered in \cite{e6ssm}.}. The VEVs of $s$ and $\overline{s}$
break $SU(3)_C \times SU(2)_W \times U(1)_Y \times U(1)_N$ to the SM gauge group. These VEVs also generate the
the following set of the mass terms in the superpotential of the model under consideration:
\be
\ba{rcl}
\delta W_{mass}&= &M^{\eta}_{ij} h_i h^c_j + M^{\zeta}_{ij} h^u_i h^d_j + M^{\xi}_{ij} s_i s_j + M^{\nu}_{ij} \nu^c_i \nu^c_j
+ \overline{M}^{\eta}\, \overline{h}\, \overline{h^c} \\[2mm]
&+& \overline{M}^{\zeta}\, \overline{h^u}\, \overline{h^d} + \overline{M}^{\xi}\, \overline{s}^2 + \overline{M}^{\nu}\, \overline{\nu^c}^2\,,
\ea
\label{ch111}
\ee
where $M^{\eta}_{ij}\sim M^{\zeta}_{ij} \sim M^{\xi}_{ij} \sim M^{\nu}_{ij} \sim \overline{M}^{\eta} \sim \overline{M}^{\zeta} \sim
\overline{M}^{\xi} \sim \overline{M}^{\nu} \sim \phi_0$ and $i,j=1,2,3$.
Two pairs of $\overline{15}$ and $15$ of $SU(6)'$ are expected to form a set of vectorlike states with masses close to $\phi_0$.
Since these supermultiplets are confined on the brane $O_{I}$, they do not interact with the superfields which reside
on the brane $O$, so that the $SU(6)\times SU(2)_N$ global symmetry of the strongly coupled sector
remains unbroken\footnote{The $SU(2)_N$ symmetry can be also broken spontaneously on the
brane $O$. Then the VEVs of $\nu^c$ and $\overline{\nu^c}$ as well as $s$ and $\overline{s}$ break the residual gauge
symmetry down to the SM gauge group, inducing the mass terms (\ref{ch111}).}.

Finally, at the scale $M_S$, which is one or two orders of magnitude lower than $M_X$, SUSY gets broken and scalar
components of all superfields including the pseudo-Goldstone bosons gain masses of order $M_S$. We assume that near
the supersymmetry breaking scale the SM singlet superfield $S$, which interacts only with the components of the $6D$
supermultiplets $\widehat{\Phi}^{u}_{3}$ and $\widehat{\Phi}^{\overline{u}}_{3}$, acquires a VEV giving rise to
the masses of the zero modes with the quantum numbers of $u_3$ and $\overline{u_3}$. Again the interactions of $S$
with other bulk $27$--plets can be forbidden by imposing the appropriate discrete $Z_2$ symmetry. This leads to the
decoupling of the right--handed top quarks from the rest of the spectrum.

The extra fermionic state $\eta$, which appears
in Eq.~(\ref{ch2}), can stem from the bulk supermultiplet that does not participate in the $E_6$ gauge interactions.
This supermultiplet can decompose under the $E_6$ gauge group and $Z_2$, $Z^{I}_2$, $Z^{II}_2$ and $Z^{III}_2$
symmetries as follows $\left(1,\,+,\,+,\,+,\,+\right)$. As a result the field content of the
weakly--coupled elementary sector given by Eq.~(\ref{ch2}) is reproduced.

\subsection{Anomaly cancellation and unification of gauge couplings}

For the consistency of the orbifold GUT model it is crucial that all anomalies get cancelled. In the $6D$ models there
are two types of anomalies: $4D$ anomalies at orbifold fixed points \cite{Adler:1969gk} and bulk anomalies
\cite{Asaka:2002my}-\cite{vonGersdorff:2006nt} which are induced by box diagrams with four gauge currents.
The contributions of the anomalous box diagrams to the $6D$ anomalies are determined by the trace of four
generators of gauge group. This trace contains a nonfactorizable part and a part which can be reduced to the product of
traces of two generators. The first part corresponds to the irreducible gauge anomaly, while the second part is known
as reducible anomaly. The reducible anomalies can be canceled by the Green--Schwarz mechanism \cite{Green:1984sg}.
On the other hand, the $6D$ orbifold GUT models based on the $E_6$ gauge group do not have an irreducible bulk anomaly
\cite{vonGersdorff:2006nt}. At the fixed points, the brane anomaly reduces to the anomaly of the
unbroken subgroup of $E_6$. It was shown that the sum of the contributions to the $4D$ anomalies at the fixed point is
equal to the sum of the contributions of the zero modes localized at the brane \cite{Asaka:2002my}, \cite{ArkaniHamed:2001is}.
In this context it is worth noting that in the orbifold GUT model under consideration the contributions of the elementary superfields,
which are confined on each brane, to the corresponding brane anomalies get cancelled automatically. Moreover
the orbifold parity assignments are chosen so that the KK modes of the bulk $27$--plets localized at the fixpoints always form
pairs of $N=1$ supermultiplets with opposite quantum numbers. This choice of parity assignments guarantees that
the contributions of zero modes of the bulk superfields to the brane anomalies are cancelled as well.

One should also mention that the orbifold GUT models do not lead to the exact gauge coupling unification at the
scale $M_X$ where $E_6$ gauge symmetry is broken. The gauge couplings at the scale $M_X$ may not be identical,
because of the sizable contributions to these couplings that can come from the branes where $E_6$ gauge symmetry is broken.
However, if in the orbifold GUT model the bulk and brane gauge couplings have almost equal strength, then
the gauge couplings, which are associated with the zero--modes of gauge bosons, are dominated by the bulk contributions
because of the spread of the wavefunction of the corresponding zero--modes. Since the bulk contributions to the
gauge couplings are necessarily $E_6$ symmetric, near the scale $M_X$ an approximate unification of the gauge couplings
is expected to take place. The gauge coupling unification within $5D$ and $6D$ orbifold GUT models was discussed in Refs.
\cite{5d-susy-ogut-proton-unif}--\cite{5d-susy-ogut-unif} and  \cite{6d-susy-ogut-unif}, respectively. As we do not require here
exact gauge coupling unification in the vicinity of the scale where $E_6$ is broken, $M_X$ can even be considerably larger than
$10^{16}\,\mbox{GeV}$, ensuring proton stability.


\section{$E_6$ inspired composite Higgs model and its phenomenological implications}

\subsection{Global symmetries and constraints}

Let us now consider the phenomenological implications of the composite Higgs model in which the weakly--coupled elementary sector
includes a set of states given by Eq.~(\ref{ch2}) at low energies, while the strongly interacting sector involves fields localised on the
brane $O$ where $E_6$ gauge symmetry is broken to $SU(6)\times SU(2)_N$. Because all fields of the strongly coupled sector reside
on the $O$ brane, the global symmetry in this sector can be $SU(6)\times SU(2)_N$ at high energies, even though local symmetry
is broken down to the SM gauge group. The $SU(3)_C\times SU(2)_W\times U(1)_Y$ gauge interactions break $SU(6)$ global symmetry.
However, if the gauge couplings of the strongly coupled sector are substantially larger than the SM gauge couplings at any scale below
$M_X$, then $SU(6)$ can still remain an approximate global symmetry of the strongly interacting sector, even at energies as low as
say $10\,\mbox{TeV}$. To simplify our analysis, we further assume that the global $SU(2)_N$ symmetry is entirely broken so that
the Lagrangian of the strongly coupled sector at low energies is just invariant under the transformations of the $SU(6)$ group only.

In this context it is worth noting that the minimal composite Higgs model (MCHM) possesses global $SO(5)$ symmetry which is broken
down to $SO(4)$ at the scale $f$ \cite{Agashe:2004rs} (for a recent review, see \cite{Bellazzini:2014yua}). The custodial symmetry
$SU(2)_{cust} \subset SO(4) \cong SU(2)_W\times SU(2)_R$ \cite{Sikivie:1980hm} allows one to protect the Peskin--Takeuchi $\hat{T}$
parameter \cite{Peskin:1991sw}, which is extremely constrained by present data \cite{Marandella:2005wd}, against new physics contributions.
Within the composite Higgs models the contributions of new states to the electroweak precision observables, including the $\hat{S}$ and $\hat{T}$
parameters as well as the $Z b_L \bar{b}_L$ coupling, were analysed in \cite{Frigerio:2011zg}, \cite{EWPOCHM}--\cite{Vignaroli:2012si}. Experimental
limits on the value of the parameter $|\hat{S}|\lesssim 0.002$ leads to the constraint $m_{\rho}=g_{\rho} f \gtrsim 2.5\,\mbox{TeV}$ where
$m_{\rho}$ is a scale associated with the masses of the set of spin-1 resonances that includes composite partners of the SM gauge bosons and
$g_{\rho}$ is a coupling of these $\rho$--like vector resonances \cite{Agashe:2004rs}.

Even more stringent bounds on $f$ come from the observed suppression of the non--diagonal flavour transitions in the case
when the matrices of effective Yukawa couplings in the strong sector, such as $Y^{u}_{ij}$ and $Y^{d}_{ij}$, are structureless, i.e anarchic matrices.
Indeed, although the generalization of the GIM mechanism in the composite Higgs model associated with partial compositeness significantly reduces
the new physics contributions to dangerous flavour--changing processes, this suppression is not sufficient to provide a fully realistic theory of flavor.
The constraints that arise from the non--diagonal flavour transitions in the quark and lepton sectors were examined
in Refs. \cite{Barbieri:2012tu}--\cite{Barbieri:2012uh} and \cite{Barbieri:2012uh}--\cite{Csaki:2008qq}, respectively.
In particular, it was shown that in the case of anarchic partial compositeness $f$ should be larger than $10\,\mbox{TeV}$, because of the constraints which stem from
the measurements of CP violation in the Kaon system \cite{Barbieri:2012tu}--\cite{Csaki:2008zd}, \cite{Redi:2011zi}--\cite{Blanke:2008zb},
as well as the measurements of the electron electric dipole moment and $\mu\to e\gamma$ transitions \cite{Agashe:2006iy}.
Large values of $f$ imply that a substantial degree of tuning is required to get a $125\,\mbox{GeV}$ Higgs state. The ratio $\xi=v^2/f^2$
constitutes a rough measure of the degree of fine--tuning and describes the departure from an elementary Higgs scenario in the composite Higgs
models. The bound on $f$ can be considerably alleviated in the composite Higgs models with flavour symmetries \cite{Barbieri:2008zt}--\cite{Barbieri:2012tu},
\cite{Redi:2011zi}, \cite{Barbieri:2012uh}--\cite{Redi:2013pga}, \cite{Cacciapaglia:2007fw}. For instance, in the models
with $U(2)^3=U(2)_{q}\times U(2)_u \times U(2)_d$ symmetry, under which the first two generations of elementary quark states transform as doublets
and the third generation as singlets, the bounds that originate from the Kaon and $B$ systems can be satisfied even for relatively low values of $f$
that correspond to $m_{\rho}\sim 3\,\mbox{TeV}$ \cite{Barbieri:2012uh}--\cite{Redi:2013pga}. Recently, the implications of the composite
Higgs models were studied for Higgs physics \cite{Bellazzini:2012tv}--\cite{Azatov:2013ura}, \cite{Mrazek:2011iu}--\cite{Pomarol:2012qf}, gauge coupling
unification \cite{Gherghetta:2004sq}, dark matter \cite{Frigerio:2011zg}--\cite{Barnard:2014tla}, \cite{Frigerio:2012uc}, \cite{Asano:2014wra} and collider
phenomenology \cite{Pomarol:2008bh}--\cite{Bellazzini:2012tv}, \cite{Barbieri:2008zt}, \cite{Redi:2011zi}, \cite{Redi:2013pga}, \cite{Pomarol:2012qf},
\cite{Delaunay:2013pwa}. The non--minimal composite Higgs models were considered in \cite{Frigerio:2011zg}--\cite{Barnard:2014tla},
\cite{Mrazek:2011iu}--\cite{Frigerio:2012uc}, \cite{Asano:2014wra}, \cite{Cacciapaglia:2014uja}.

The composite Higgs model under consideration (E$_6$CHM) does not possess $SU(2)_{cust}$ symmetry mentioned above.
As a consequence, the absolute value of the parameter $|\hat{T}|$ is expected to be of the order \cite{Frigerio:2011zg}
\be
|\hat{T}|\sim \xi=\frac{v^2}{f^2}\,.
\label{ch12}
\ee
Since the electroweak precision measurements constrain $|\hat{T}|\lesssim 0.002$, Eq.~(\ref{ch12}) leads to the stringent lower bound
on the scale $f\gtrsim 5-6\,\mbox{TeV}$\footnote{A weaker bound was obtained in \cite{Cheng:2013qwa}.}, where the breakdown of the
$SU(6)$ global symmetry takes place. Although the adequate suppression of the flavour--changing processes in general requires $f$
to be even larger, i.e. $f\gtrsim 10\,\mbox{TeV}$, the desirable suppression of the non--diagonal flavour transitions can also be achieved
by imposing $U(2)^3$ or even larger flavour symmetry, just as in other composite Higgs models discussed above. Therefore, hereafter we
assume that $f\gtrsim 5-10\,\mbox{TeV}$. This means that a significant tuning, $\sim 0.1-0.01\%$, is needed to comply with the Higgs mass
measurements. This tuning can be accomplished by cancelling two different contributions associated with the exotic fermions and gauge
fields that appear with different signs \cite{Barnard:2014tla}.

In contrast to the SM where there are two accidental $U(1)$ symmetries ($U(1)_B$ and $U(1)_L$) of the renormalizable Lagrangian that
result in the conservation of baryon and lepton numbers, new interactions in the composite Higgs models in general give rise to
baryon and lepton number violating processes. Indeed, because of the mixing between the elementary states and their composite partners,
the four-fermion operators leading to proton decay can be generated through non--perturbative effects. Such operators are only
suppressed by the scale $f$ and the small fractions of compositeness of the first and second generation fermions. This suppression is not
sufficient to prevent too rapid proton decay and other baryon number violating processes. Similarly, dimension-5 operators of the
form $\ell_i \ell_j H H/ f$, where $H$ is a composite Higgs doublet, can be induced resulting in lepton number violation and generating
Majorana neutrino masses which are far too large with respect to the observed ones.

Thus in the composite Higgs models one is forced to impose additional $U(1)_B$ and $U(1)_L$ symmetries to avoid violations of
baryon and lepton numbers which are too large. These symmetries should be part of the global symmetries of the composite sector and should be extended
consistently to the elementary sector so that baryon and lepton numbers are preserved  to very good approximation up to scales $\sim M_X$.
In particular, within the E$_6$CHM the interactions between the elementary states and their composite partners break $SU(6)$ global
symmetry and its $SU(5)$ subgroup but must preserve its $SU(3)_C\times SU(2)_W\times U(1)_Y$ gauged subgroup as well as global
$U(1)_B$ and $U(1)_L$ symmetries. For this reason, the simplest possibility is to take $U(1)_B\times U(1)_L$ external to the group $SU(6)$,
because $SU(6)$ and its $SU(5)$ subgroup always allow for baryon and lepton number violating operators in the elementary sector.
In other words, at low energies the Lagrangian of the strongly coupled sector of the E$_6$CHM should be invariant under the transformations of
an $SU(6)\times U(1)_B\times U(1)_L$ global symmetry, whereas the full effective Lagrangian of the E$_6$CHM respects
$SU(3)_C\times SU(2)_W\times U(1)_Y \times U(1)_B\times U(1)_L$ symmetry.

The $U(1)_B$ and $U(1)_L$ symmetries can be incorporated into the orbifold GUT model considered in the previous section.
Since this model is based on the $E_6\times G_0$ gauge symmetry, $U(1)_B$ and $U(1)_L$ can be subgroups of the $G_0$ group
associated with the strongly coupled sector. In principle the $G_0$ symmetry can be broken down to its subgroup $G$ in such a way that
the $U(1)_B$ and $U(1)_L$ symmetries remain intact on the brane $O$, where all composite sector fields reside. As a result the Lagrangian
of the strongly coupled sector respects the $SU(6)\times U(1)_B\times U(1)_L$ global symmetry.

Nevertheless, both $U(1)_B$ and $U(1)_L$ symmetries are expected to get broken on the brane $O_{I}$. Note that the
nearly exact conservation of the $U(1)_B$ and $U(1)_L$ charges at low energies implies that the elementary fermions with different
baryon and/or lepton numbers should belong to different bulk $27$--plets. In this sense the baryon and lepton numbers of the bulk
supermultiplets are determined by the $U(1)_B$ and $U(1)_L$ charges ($B$ and $L$) of the fermion components of these
supermultiplets that survive to low energies. Thus $\widehat{\Phi}^{u}_{i}$ and $\widehat{\Phi}^{d}_{i}$ have $B=-\dfrac{1}{3}$
and $L=0$, $\widehat{\Phi}^{q}_{i}$ carry $B=\dfrac{1}{3}$ and $L=0$, $\widehat{\Phi}^{\ell}_{i}$ have $B=0$ and $L=1$
whereas $\widehat{\Phi}^{e}_{i}$ carry $B=0$ and $L=-1$. Because all components of the
bulk supermultiplets carry the same $U(1)_B$ and $U(1)_L$ charges, $E_6$ gauge interactions do not give rise to baryon and
lepton number violating operators, in contrast with conventional GUTs. On the other hand, the breakdown of $U(1)_B$ occurs
when the zero modes of the components $h^c$ and $h$ of the bulk supermultiplets $\widehat{\Phi}^{u}_{i}$ and $\widehat{\Phi}^{d}_{i}$
form vectorlike states. The corresponding mass terms in the Lagrangian are forbidden by the $U(1)_B$ symmetry. The Majorana mass
terms associated with the zero modes of the $\nu^c$ and $s$ components of the bulk supermultiplets $\widehat{\Phi}^{d}_{i}$
also break this symmetry. Finally, the breakdown of both $U(1)_B$ and $U(1)_L$ symmetries takes place when the zero modes of the
components $h^u$ and $h^d$ of the bulk $27$--plets $\widehat{\Phi}^{q}_{i}$ and $\widehat{\Phi}^{\ell}_{i}$ form vectorlike states.
All these mass terms are induced by the VEVs of the scalar components of the superfields which are localised on the brane $O_{I}$.
Therefore $U(1)_B$ and $U(1)_L$ should be broken on this brane. Then the phenomenological viability of the model under consideration
requires that the masses of the elementary states, which cause the breakdown of $U(1)_B$ and $U(1)_L$ symmetries, should be
sufficiently close to $10^{16}\,\mbox{GeV}$ to ensure the adequate suppression of the baryon and lepton number violating operators
that give rise to proton decay. In the context of $5D$ and $6D$ orbifold GUT models, proton stability was discussed in Refs.
\cite{5d-susy-ogut-proton}--\cite{5d-susy-ogut-proton-unif} and \cite{Buchmuller:2004eg}, respectively.

\subsection{Non--linear realization of the Higgs mechanism}

Below scale $f$ ($f\gtrsim 5-10\,\mbox{TeV}$) the global $SU(6)$ symmetry in the E$_6$CHM is broken down to $SU(5)$,
which in turn contains the $SU(3)_C\times SU(2)_W\times U(1)_Y$
subgroup. Here we denote the unbroken generators of $SU(6)$, i.e. generators of its $SU(5)$ subgroup, by $T^a$, while the broken ones,
i.e. generators from the coset $SU(6)/SU(5)$, are denoted by $T^{\hat{a}}$. The generators of the $SU(6)$ group are normalised here so
that $\mbox{Tr} T^a T^b = \ds\frac{1}{2} \delta_{ab}$. There are eleven pNGB states in the $SU(6)/SU(5)$ coset space.
These can be parameterised by
\be
\Sigma= e^{i\Pi/f}\,,\qquad \Pi=\Pi^{\hat{a}} T^{\hat{a}}\,,
\label{ch13}
\ee
where $f$ plays the role of a decay constant. The matrix $\Pi$ is given by
\be
\Pi=\left(
\begin{array}{cccccc}
-\ds\frac{\phi_0}{\sqrt{60}} & 0 & 0 & 0 & 0 & \ds\frac{\phi_1}{\sqrt{2}} \\
0 & -\ds\frac{\phi_0}{\sqrt{60}} & 0 & 0 & 0 & \ds\frac{\phi_2}{\sqrt{2}} \\
0 & 0 & -\ds\frac{\phi_0}{\sqrt{60}} & 0 & 0 & \ds\frac{\phi_3}{\sqrt{2}} \\
0 & 0 & 0 & -\ds\frac{\phi_0}{\sqrt{60}} & 0 & \ds\frac{\phi_4}{\sqrt{2}} \\
0 & 0 & 0 & 0 & -\ds\frac{\phi_0}{\sqrt{60}} & \ds\frac{\phi_5}{\sqrt{2}} \\
\ds\frac{\phi^{\dagger}_1}{\sqrt{2}} & \ds\frac{\phi^{\dagger}_2}{\sqrt{2}} & \ds\frac{\phi^{\dagger}_3}{\sqrt{2}} &
\ds\frac{\phi^{\dagger}_4}{\sqrt{2}} & \ds\frac{\phi^{\dagger}_5}{\sqrt{2}} & \ds\frac{5 \phi_0}{\sqrt{60}}
\end{array}
\right)\,.
\label{ch14}
\ee

To write the non--linear realization of the Higgs mechanism in the E$_6$CHM, it is convenient to
choose a specific direction for the vacuum $\Omega_0$. In particular, the breaking $SU(6)\to SU(5)$ can be
parameterised through the fundamental representation of $SU(6)$, i.e.
\be
\Omega_0^T= (0\quad 0\quad 0\quad 0\quad 0\quad 1)\,.
\label{ch15}
\ee
Then the leading order Lagrangian that describes the interactions of the pNGB states can be written as
\be
\mathcal{L}_{pNGB}=\ds\dfrac{f^2}{2}\biggl|\mathcal{D}_{\mu} \Omega \biggr|^2\,.
\label{ch16}
\ee
In Eq.~(\ref{ch16}) the non--linear representation of the pNGB states is obtained in terms of
a 6--component unit vector $\Omega$ that reads
\be
\Omega=\Sigma\, \Omega_0,\quad \Omega^T=e^{i\frac{\phi_0}{\sqrt{15}f}}
\Biggl(C \phi_1\quad C \phi_2\quad C \phi_3\quad C \phi_4\quad C\phi_5\quad \cos\dfrac{\tilde{\phi}}{\sqrt{2} f} + \sqrt{\dfrac{3}{10}} C \phi_0 \Biggr),
\label{ch17}
\ee
where
$$
C=\dfrac{i}{\tilde{\phi}} \sin \dfrac{\tilde{\phi}}{\sqrt{2} f}\,,\qquad \tilde{\phi}=\sqrt{\dfrac{3}{10}\phi_0^2+\phi_1^2+\phi_2^2+\phi_3^2+\phi_4^2+\phi_5^2}\,.
$$
Since $\tilde{\phi}$ and $\phi_0$ are invariant under the preserved $SU(5)$, $\Omega$ transforms as $\bf{5}+\bf{1}$ under the transformation of
the $SU(5)$ group. Therefore one can introduce a 5--component vector, $\tilde{H}\sim (\phi_1\, \phi_2\, \phi_3\, \phi_4\, \phi_5)$. The first two components of
this vector transform as an $SU(2)_W$ doublet, $H\sim (\phi_1\, \phi_2)$, and therefore $H$ is associated with the SM--like Higgs doublet. Three other
components, $T\sim (\phi_3\, \phi_4\, \phi_5)$, correspond to an $SU(3)_C$ triplet. Because in the SM the Higgs doublet has $B=L=0$, no components of $\Omega$
should carry any baryon and/or lepton numbers.

The low energy effective Lagrangian of the E$_6$CHM, that includes the interactions among the SM fields, the pNGB states and exotic fermions,
can be obtained by integrating out the heavy resonances of the composite sector. However, only interactions, that break global $SU(6)$ symmetry, can
induce the pNGB effective potential $V_{eff}(\tilde{H}, T, \phi_0)$ which must vanish in the exact $SU(6)$ symmetry limit. As a consequence, the main
contributions to $V_{eff}(H, T, \phi_0)$ should come from the interactions of the elementary fermions and gauge bosons with their composite partners
which explicitly break $SU(6)$ symmetry. The analysis of the structure of the pNGB effective potential within similar composite Higgs models,
including the derivation of quadratic terms $m^2_H |H|^2$ and $m_T^2 |T|^2$, shows that there is a substantial part of the parameter space
where $m_H^2$ tends to be negative while $m_T^2$ remains positive \cite{Frigerio:2011zg}--\cite{Barnard:2014tla}. In other words, in this parameter
region EW symmetry is broken, whereas $SU(3)_C$ colour is preserved. This happens when the contributions of the top quark and exotic fermions to $m_H^2$
are negative and sufficiently large to overcome the gauge boson contribution to $m_H^2$. At the same time, in this case $m_T^2$ can be positive due to
the large contribution to $m_T^2$ generated by the interactions of gluons and their composite partners. Therefore, hereafter we just assume that
the non--zero components of the vector $\Omega$ break $SU(6)$ symmetry so that $SU(2)_W\times U(1)_Y$ gauge symmetry gets broken down to
$U(1)_{em}$, associated with electromagnetism, whereas $SU(3)_C$ symmetry remains intact.

\subsection{Generation of masses of the SM fermions}

As mentioned before, all elementary quark and lepton states gain masses through the mixing with their composite partners.
Thus it is important to ensure that the corresponding mixing can occur within the E$_6$CHM. In the model under consideration
different multiplets of elementary quarks and leptons stem from different representations of the $SU(6)$ subgroup of $E_6$.
All other components of the corresponding $SU(6)$ representations are extremely heavy (see Section 2). Thus, at low energies
elementary quarks and leptons appear as incomplete multiplets of $SU(6)$, which decompose under
the $SU(6)\times U(1)_B\times U(1)_L$ global symmetry as follows
\be
\begin{array}{c}
u^c_{\alpha} \in {\bf 15}^u_{\alpha}=\left({\bf 15},\,-\dfrac{1}{3},\,0\right)_{\alpha}\quad
q_i \in {\bf 15}^q_i=\left({\bf 15},\,\dfrac{1}{3},\,0\right)_{i} \quad
d^c_i \in {\bf \overline{6}}^d_i=\left({\bf \overline{6}},\,-\dfrac{1}{3},\,0\right)_{i} \\[3mm]
e^c_i \in {\bf 15}^e_i=\Biggl({\bf 15},\,0,\,-1\Biggr)_{i}\qquad
\ell_i \in {\bf \overline{6}}^{\ell}_i=\Biggl({\bf \overline{6}},\,0,\,1 \Biggr)_{i}\,,
\end{array}
\label{ch18}
\ee
where the first, second and third quantities in brackets are the $SU(6)$ representation, $U(1)_{B}$ and $U(1)_{L}$ charges,
respectively, while $\alpha=1,2$ and $i=1,2,3$. The composite partners of the elementary fermions should be embedded into the
$SU(6)$ representations so that all quark and lepton Yukawa interactions of the SM, which induce non--zero fermion masses,
are allowed.

In our analysis we use the simplest $SU(5)$ GUT as a guideline. Below scale $f$, where $SU(6)$ global symmetry is broken
down to $SU(5)$, the elementary quark and lepton states constitute the following incomplete $SU(5)$ multiplets
\be
\begin{array}{c}
u^c_{\alpha} \in {\bf 10}^u_{\alpha}=\left({\bf 10},\,-\dfrac{1}{3},\,0\right)_{\alpha}\quad
q_i \in {\bf 10}^q_i=\left({\bf 10},\,\dfrac{1}{3},\,0\right)_{i} \quad
d^c_i \in {\bf \overline{5}}^d_i=\left({\bf \overline{5}},\,-\dfrac{1}{3},\,0\right)_{i} \\[3mm]
e^c_i \in {\bf 10}^e_i=\Biggl({\bf 10},\,0,\,-1\Biggr)_{i}\qquad
\ell_i \in {\bf \overline{5}}^{\ell}_i=\Biggl({\bf \overline{5}},\,0,\,1 \Biggr)_{i}\,,
\end{array}
\label{ch19}
\ee
where, as before, the second and third quantities in brackets correspond to the baryon and lepton numbers of
these $SU(5)$ representations. If the particle content of this model involved an elementary Higgs boson, then the Higgs doublet
$h$ could be embedded into the fundamental representation of $SU(5)$, i.e. $h \in {\bf 5}^{h}=\left({\bf 5},\,0,\,0\right)$.
In this case the Yukawa couplings of the up type quarks have the following $SU(5)$ structure
\be
\mathcal{L}^u_{SU(5)}\simeq h^u_{\alpha i}\, {\bf 10}^u_{\alpha}\, {\bf 10}^q_i\, {\bf 5}^{h}.
\label{ch20}
\ee

In order to reproduce the SM up--quark Yukawa couplings, one is forced to assume that interactions similar to
those given by Eq.~(\ref{ch20}) are reproduced in the strongly coupled sector below the scale $f$. In the case of
the $SU(6)$ symmetry, the Higgs multiplet ${\bf 5}^{h}$ should be replaced by the unit vector $\Omega$. Instead of
two other $SU(5)$ representations ${\bf 10}^u_{\alpha}$ and ${\bf 10}^q_i$, that appear in Eq.~(\ref{ch20}), one
should include two $SU(6)$ multiplets that contain an $SU(5)$ decuplet. The simplest $SU(6)$ representation of this
type is an antisymmetric second--rank tensor field ${\bf 15}$. The next--to--simplest $SU(6)$ representation, that
involves an $SU(5)$ decuplet, is a totally antisymmetric third--rank tensor ${\bf 20}$. These $SU(6)$ representations
have the following decomposition in terms of $SU(5)$ representations: ${\bf 15}={\bf 10} \oplus {\bf 5}$ and
${\bf 20}={\bf 10} \oplus {\bf \overline{10}}$. The presence of the ${\bf 15}$--plet and ${\bf 20}$--plet allows
for the generalisation of the $SU(5)$ structure of the up--quark Yukawa interactions (\ref{ch20}) to the case of
$SU(6)$ symmetry that results in
\be
\mathcal{L}^u_{SU(6)}\sim {\bf 20}\times {\bf 15}\times {\bf 6}\,,
\label{ch21}
\ee
where ${\bf 6}$ should be identified with the unit vector $\Omega$.

The structure of the interactions (\ref{ch21}) leads to two different scenarios. In scenario A, the composite partners
of $u^c_{\alpha}$ and $q_i$ ($U_{\alpha}$ and $Q_i$) belong to ${\bf 15}(U_{\alpha})$ and ${\bf 20}(Q_i)$
representations of $SU(6)$, whereas in scenario B the composite partners of $u^c_{\alpha}$ and $q_i$ are
components of ${\bf 20}(U_{\alpha})$ and ${\bf 15}(Q_i)$, respectively. In principle, the $SU(6)$ symmetry forbids
the mixing between the components of ${\bf 20}$--plets, that contain the composite partners of quarks, and
${\bf 15}$--plets which involve the elementary quark states. Nevertheless, such mixing can be induced below
the scale $f$, where the $SU(6)$ global symmetry is broken down to $SU(5)$. To demonstrate this, let us focus on
scenario A and assume that the strongly interacting sector includes not only ${\bf 20}(Q_i)$ but also ${\bf 15}(Q'_i)$
and ${\bf \overline{15}}(\overline{Q'}_i)$. Then the part of the Lagrangian that determines the mixing between
the elementary quark states $q_i$ and their composite partners $Q_i$ can be written as
\be
\begin{array}{c}
\mathcal{L}^q_{mix}=\sigma_Q f\, {\bf 20}(Q_i) {\bf 15}(Q'_i) \Omega + m_{Q} {\bf 20}(Q_i) {\bf 20}(Q_i)\qquad\qquad\qquad\qquad\qquad\\[2mm]
\qquad\qquad\qquad\qquad\qquad+ m_{Q'} {\bf 15}(Q'_i) {\bf \overline{15}}(\overline{Q'}_i)
+ \mu_{q} {\bf \overline{15}}(\overline{Q'}_i) {\bf 15}^q_i\, .
\end{array}
\label{ch22}
\ee
When $\sigma_Q f \gg m_{Q}\sim m_{Q'} \gg \mu_{q}$, the ${\bf \overline{10}}$ from ${\bf 20}(Q_i)$ and the ${\bf 10}$ from ${\bf 15}(Q'_i)$
form heavy vector-like states with masses $\sim \sigma_Q f$, so that these states are almost decoupled
from the rest of the particle spectrum. The remaining ${\bf 10}$ from ${\bf 20}(Q_i)$ and the components of ${\bf 10}^q_i$ get mixed.
If $\dfrac{m_{Q} m_{Q'}}{\sigma_Q f}\gg \mu_{q}$, then one superposition of these ${\bf 10}$--plets, which is
predominantly ${\bf 10}$ from ${\bf 20}(Q_i)$, and ${\bf \overline{10}}$--plet from ${\bf \overline{15}}(\overline{Q'}_i)$
get combined, forming vector-like states that acquire masses of order of $\dfrac{m_{Q} m_{Q'}}{\sigma_Q f}$.
Another superposition of ${\bf 10}$ from ${\bf 20}(Q_i)$ and ${\bf 10}^q_i$, which is basically a superposition  of elementary quark
states $q_i$ and their composite partners, gain masses after the EW symmetry breaking. Similarly, the mixing between the
components of the incomplete ${\bf 15}^u_{\alpha}$ and their composite partners from ${\bf 20}(U_{\alpha})$ can be induced
in the case of the scenario B.

In the $SU(5)$ models the masses of the down type quarks are induced through the Yukawa interactions
\be
\mathcal{L}^d_{SU(5)}\simeq h^d_{i j}\, {\bf 10}^q_i\, {\bf \overline{5}^d_j}\, {\bf \overline{5}}^{h}\,.
\label{ch23}
\ee
The simplest $SU(6)$ generalisation of the $SU(5)$ structure of the down--quark Yukawa interactions (\ref{ch23})
takes the form:
\be
\mathcal{L}^d_{SU(6)}\sim {\bf 15}\times {\bf \overline{6}}\times {\bf \overline{6}'}\,.
\label{ch24}
\ee
In scenario B the Yukawa couplings (\ref{ch24}) can be used to generate the masses of the down type quarks
after EW symmetry breaking. In this case the ${\bf \overline{6}'}$  in Eq.~(\ref{ch24}) has to be identified with
$\Omega^{\dagger}$, the ${\bf 15}$ should be associated with ${\bf 15}(Q_i)$ and the ${\bf \overline{6}}$ is expected
to contain the composite partners of $d^c_i$ ($D_i$), i.e. ${\bf \overline{6}}\equiv {\bf \overline{6}}(D_i)$.
The $SU(6)$ symmetry does not forbid mixing between ${\bf 15}^q_i$ and ${\bf 15}(Q_i)$ or
between ${\bf \overline{6}}^d_i$ and  ${\bf \overline{6}}(D_i)$.

In the case of scenario A, the simplest $SU(6)$ generalisation of the Yukawa interactions (\ref{ch23}),
that can give rise to the non--zero masses of the SM down type quarks, is given by
\be
\mathcal{L}^d_{SU(6)}\sim {\bf 20}\times {\bf \overline{15}}\times {\bf \overline{6}'}\,,
\label{ch25}
\ee
where again ${\bf \overline{6}'}\equiv \Omega^{\dagger}$, while the ${\bf 20}$--plet corresponds to the $SU(6)$
representations that involve composite partners of $q_i$, i.e. ${\bf 20}(Q_i)$, and the ${\bf \overline{15}}$--plet
should contain composite partners of $d^c_i$, i.e. ${\bf \overline{15}}\equiv {\bf \overline{15}}(D_i)$.
As pointed out earlier, the mixing between components of the incomplete ${\bf 15}^q_i$ multiplets and
composite partners of $q_i$ from ${\bf 20}(Q_i)$ can be induced below scale $f$. The breakdown of $SU(6)$
symmetry can also give rise to the mixing between the corresponding components of the incomplete
${\bf \overline{6}}^d_i$ multiplet and ${\bf \overline{15}}(D_i)$. This happens, for example, when the
strongly coupled sector involves ${\bf \overline{15}}(D_i)$ and ${\bf 15}(\overline{D}_i)$ as well as
${\bf \overline{6}}(D'_i)$ and ${\bf 6}(\overline{D}'_i)$. The part of the Lagrangian that leads to the
mixing of the elementary quark states, $d^c_i$, and their composite partners, $D_i$, can be written in
the following form:
\be
\begin{array}{c}
\mathcal{L}^d_{mix}=m_{D} {\bf \overline{15}}(D_i) {\bf 15}(\overline{D}_i) +
\sigma_d f\, {\bf 15}(\overline{D}_i) {\bf \overline{6}}(D'_i) \Omega^{\dagger} \qquad\qquad\qquad\qquad\qquad\\[2mm]
\qquad\qquad\qquad\qquad\qquad+ m_{D'} {\bf \overline{6}}(D'_i) {\bf 6}(\overline{D}'_i)
+ \mu_{d} {\bf 6}(\overline{D}'_i) {\bf \overline{6}}^d_i \, .
\end{array}
\label{ch26}
\ee
If $m_{D'}\gg m_{D},\, \sigma_d f$ and $\mu_d$ the composite states ${\bf \overline{6}}(D'_i)$ and
${\bf 6}(\overline{D}'_i)$ can be integrated out.  Then the second term in the Lagrangian (\ref{ch26}) results
in mixing between the components of the incomplete ${\bf \overline{6}}^d_i$ multiplet and their
composite partners from ${\bf \overline{15}}(D_i)$.

Since charged lepton and down--quark Yukawa interactions have the same $SU(5)$ structure in the simplest
$SU(5)$ GUT, the $SU(6)$ generalisation of these interactions (\ref{ch24}) can be used in both scenarios A and B
to generate the masses of charged leptons within the E$_6$CHM. Again one can set
${\bf \overline{6}'}\equiv \Omega^{\dagger}$ in Eq.~(\ref{ch24}). At the same time one can expect that
the composite partners of $e^c_i$ and $\ell_i$ are components of ${\bf 15}(E_i)$ and ${\bf \overline{6}}(L_i)$,
respectively. The mixing between the components of ${\bf 15}^e_i$ and their composite partners from
${\bf 15}(E_i)$, as well as the mixing between the corresponding components of ${\bf \overline{6}}^{\ell}_i$
and ${\bf \overline{6}}(L_i)$, are not forbidden by the $SU(6)$ symmetry. Therefore such Yukawa interactions
should lead to non--zero masses for the charged leptons after the breakdown of the EW symmetry.

The masses of the elementary left--handed neutrinos in the $SU(5)$ GUT are induced through the Yukawa
interactions
\be
\mathcal{L}^{\nu}_{SU(5)}\simeq h^{\nu}_{i j}\, {\bf \overline{5}^{\ell}_i}\, {\bf 5}^{h} {\bf 1}_j\,,
\label{ch27}
\ee
where ${\bf 1}_i$ correspond to Majorana right--handed neutrinos that do not participate in the SM gauge
interactions. The simplest $SU(6)$ generalisation of the Yukawa couplings (\ref{ch27}) is given by
\be
\mathcal{L}^{\nu}_{SU(6)}\sim {\bf \overline{6}}\times {\bf 6'}\times {\bf 1}\,.
\label{ch28}
\ee
In Eq.~(\ref{ch28}) ${\bf \overline{6}}' \equiv \Omega$ and the ${\bf \overline{6}}$ should be associated
with ${\bf \overline{6}}(L_i)$. The Yukawa interactions (\ref{ch28}) imply that the dynamics of the strongly
coupled sector should lead to the formation of a set of the $SU(6)$ singlet bound states with spin $1/2$.
Because in the composite sector $U(1)_L$ symmetry is preserved, these fermion bound states $N_i$ and
$\overline{N}_i$ have to carry lepton number, so that $N_i=\Biggl({\bf 1}, \,0,\,-1 \Biggr)$ and
$\overline{N}_i=\Biggl({\bf 1}, \,0,\,1 \Biggr)$. To ensure the smallness of the masses of the
elementary left--handed neutrinos one can include in the elementary sector a set of heavy Majorana
states, $S_i$, that get mixed with $\overline{N}_i=\Biggl({\bf 1}, \,0,\,1 \Biggr)$. Such fermionic states
may come from the bulk supermultiplets that do not participate in the $E_6$ gauge interactions but
carry lepton number. The Majorana masses of $S_i$ can be generated after the breakdown of the $U(1)_L$
symmetry on the brane $O_{I}$. These masses can be somewhat lower than $M_X$.

In the case of one lepton flavour the simplest low--energy effective Lagrangian of the type discussed above
can be written as
\be
\begin{array}{c}
\mathcal{L}^{e\nu}_{eff} = \mu_e (\overline{E} e^c) + \mu_{\ell} (\overline{L} \ell) +
M_E (\overline{E} E) + M_L (\overline{L} L) + h_E (L H^{\dagger}) E + h_N (L H) N \\[2mm]
+ M_N (N \overline{N}) + \mu_N (\overline{N} S) + M_S (SS)+ h.c.\,,
\end{array}
\label{ch29}
\ee
where $E$, $\overline{E}$, $L$, $\overline{L}$, $N$ and $\overline{N}$ are composite fermions, while $H$ is a composite
Higgs doublet. In the limit where lepton number is conserved, i.e. the parameter $\mu_N$ vanishes, the Lagrangian (\ref{ch29})
results in a massless Majorana fermion that can be identified with the elementary left--handed neutrino if $\mu_{\ell}\to 0$.
Assuming that the mixing between the elementary and composite states is rather small, i.e.
$\mu_e,\,\mu_{\ell},\,\mu_N \ll M_{E},\,M_{L},\, M_{N}$, one can obtain the approximate expressions for the
masses of the elementary charged lepton and left--handed neutrino states ($m_e$ and $m_{\nu}$)
\be
|m_e| \simeq h_E \left( \dfrac{\mu_e}{M_E} \right) \left( \dfrac{\mu_{\ell}}{M_L} \right) \dfrac{v}{\sqrt{2}},\qquad
|m_{\nu}| \simeq h^2_{N} \left( \dfrac{\mu_{\ell}}{M_L} \right)^2 \left( \dfrac{\mu_{N}}{M_N} \right)^2 \dfrac{v^2}{2 M_S}\,.
\label{ch30}
\ee
From Eq.~(\ref{ch30}) it follows that $|m_{\nu}|\ll |m_e|$ if the mixing between the elementary and composite states
is small and/or $M_S\gg v$.

\subsection{Implications for collider phenomenology and dark matter}

As pointed out in the introduction to this article, the composite Higgs model under consideration implies
that in the exact $SU(6)$ symmetry limit the dynamics of the strongly interacting sector gives rise to massless
$SU(6)$ representations that contain composite $t^c$. In scenario A the right--handed top quark state
belongs to a ${\bf{15}}$--plet that must carry the same baryon number as $t^c$, i.e. $B_{15}=-1/3$. In this
case we assume that in addition to the ${\bf{15}}$--plet, two ${\bf \overline{6}}$--plets (${\bf \overline{6}}_1$
and ${\bf \overline{6}}_2$) with spin $1/2$ and opposite baryon numbers remain massless as well that leads
to the $SU(6)$ anomaly cancellation in the massless sector. Moreover, we allow for interaction between vector
$\Omega$ and multiplets ${\bf{15}}$ and ${\bf \overline{6}}_1$ of the type
${\bf{15}}\times {\bf \overline{6}}_1 \times \Omega^{\dagger}$ that do not violate the $U(1)_B$ symmetry if
$B_{\bar{6}_1}=-B_{15}=1/3$. Such a Yukawa coupling results in the formation of vector--like states that
involve a ${\bf 5}$--plet from ${\bf 15}$ and a ${\bf \overline{5}}$--plet from ${\bf \overline{6}}_1$. The $SU(5)$
singlet components of ${\bf \overline{6}}_1$ and ${\bf \overline{6}}_2$ can also acquire mass through the
interaction $({\bf \overline{6}}_1 \Omega)(\Omega {\bf \overline{6}}_2)$. As a consequence, only the ${\bf 10}$--plet
from ${\bf 15}$ and ${\bf \overline{5}}$--plet from ${\bf \overline{6}}_2$, that carry $B=-1/3$, do not acquire
masses by interacting with $\Omega$. Nevertheless, these ${\bf 10}$--plet and ${\bf \overline{5}}$--plet states get
combined with elementary exotic states $\bar{q}$, $\bar{d^c}$, $\bar{\ell}$, $\bar{e^c}$, resulting in a
set of vector--like states with masses somewhat below $f$ and a composite right--handed top quark.

In scenario B the right--handed top quark state belongs to the ${\bf{20}}$--plet of $SU(6)$ with
baryon number $B_{20}=-1/3$. We assume that in this case the dynamics of the strongly coupled sector results
in a massless ${\bf{20}}$--plet, ${\bf{15}}$--plet (${\bf 15}'$) and two ${\bf \overline{6}}$--plets
(${\bf \overline{6}}'_1$ and ${\bf \overline{6}}'_2$) with spin $1/2$ in the exact $SU(6)$ symmetry limit.
The interaction between the ${\bf{20}}$--plet, ${\bf 15}'$ and vector $\Omega$ of the type
${\bf{20}}\times {\bf 15}' \times \Omega$ gives rise to the formation of vector--like states that
involve the ${\bf \overline{10}}$--plet from ${\bf 20}$ and the ${\bf 10}$--plet from ${\bf 15}'$, whereas the
coupling of ${\bf 15}'$ to ${\bf \overline{6}}'_1$ and $\Omega$, i.e.
${\bf{15}}' \times {\bf \overline{6}}'_1 \times \Omega^{\dagger}$, leads to the massive states
composed of a ${\bf 5}$--plet from ${\bf 15}'$ and a ${\bf \overline{5}}$--plet from ${\bf \overline{6}}'_1$.
Again the mass of the $SU(5)$ singlet components of ${\bf \overline{6}}'_1$ and ${\bf \overline{6}}'_2$
can be induced through the interaction $({\bf \overline{6}}'_1 \Omega)(\Omega {\bf \overline{6}}'_2)$.
None of these interactions are forbidden by the $U(1)_B$ symmetry provided
$B_{20}=-B_{15'}=B_{\overline{6}'_1}=-B_{\overline{6}'_2}=-1/3$. As before one ${\bf 10}$--plet
from ${\bf{20}}$ and one ${\bf \overline{5}}$--plet from ${\bf \overline{6}}'_2$, that do not gain masses
because of the interaction with vector $\Omega$, as well as elementary exotic states $\bar{q}$, $\bar{d^c}$,
$\bar{\ell}$, $\bar{e^c}$ form a set of vector--like states and composite $t^c$. However, in contrast to
scenario A, the ${\bf 10}$--plet from ${\bf{20}}$ and the ${\bf \overline{5}}$--plet from ${\bf \overline{6}}'_2$
have opposite baryon numbers, $-1/3$ and $1/3$ respectively.

Thus in both scenarios the set of vector--like fermion states, that can have masses in the few TeV range, include:
colour triplet $t' (\overline{t}')$ with electric charges $+ 2/3 (-2/3)$; colour triplets of  quarks $b'_1$ and $b'_2$
($\overline{b}'_1$ and $\overline{b}'_2$) with different masses but the same electric charge $- 1/3 (+1/3)$;
colourless fermions $e'_1$ and $e'_2$ ($\overline{e}'_1$ and $\overline{e}'_2$) with different masses but the
same electric charge $- 1 (+1)$;  as well as a neutral fermion state $\nu'$ $(\overline{\nu}')$ which is formed
by the components of the $SU(2)_W$ doublets. Baryon number conservation implies that all these fermion
states carry non--zero $U(1)_B$ charges. In both cases $t'$, $b'_1$ and $\overline{e}'_1$ have baryon number
$-1/3$. In scenario A $e'_2$, $\nu'$ and $\overline{b}'_2$ carry baryon number $-1/3$, while in scenario B
these states have opposite baryon number, $+1/3$. The set of the lightest exotic states should be supplemented
by the scalar colour triplet $T$ ($T^{\dagger}$) with electric charge $-1/3$ ($+1/3$) and zero baryon number,
that stem from the pNGB $5$--plet, $\tilde{H}$, that also gives rise to the Higgs doublet, $H$.

One of the lightest exotic states in the E$_6$CHM should be stable. This can be understood in terms of the
$Z_3$ symmetry which is known as baryon triality (see, for example \cite{Frigerio:2011zg}, \cite{baryon-triality}).
The corresponding transformations can be defined as
\be
\Psi \longrightarrow e^{2\pi i B_3/3} \Psi,\qquad B_3 = (3 B - n_C)_{\mbox{mod}\,\, 3}\,,
\label{ch31}
\ee
where $B$ is the baryon number of the given multiplet $\Psi$ and $n_C$ is the number of colour indices ($n_C=1$ for the
colour triplet and $n_C=-1$ for ${\bf \overline{3}}$). Because baryon number is preserved  to a very good approximation,
the low energy effective Lagrangian of the E$_6$CHM is invariant under the transformations of this discrete
$Z_3$ symmetry. All bosons and fermions in the SM have $B_3=0$. On the other hand, in both scenarios $B_3(T)=2$ and
$B_3(t')=B_3(b'_1)=B_3(e'_1)=1$. At the same time, in scenario A $B_3(b'_2)=0$ and $B_3(e'_2)=B_3(\nu')=2$,
whereas in scenario B $B_3(b'_2)=B_3(e'_2)=B_3(\nu')=1$. Because of the invariance of the low energy effective
Lagrangian of the E$_6$CHM with respect to the transformations of baryon triality, the lightest exotic state
with non--zero $B_3$ charge can not decay into SM particles and must therefore be stable. The decay of such an exotic
state can be induced by baryon number violating operators and is therefore extremely strongly suppressed.

If the lightest states with non--zero $B_3$ charge are exotic colour triplets or exotic charged fermions then these states would
have been copiously produced during the very early epochs of the Big Bang. Those strong or electromagnetically
interacting lightest exotic states which survive annihilation would subsequently have been confined in heavy hadrons
which would annihilate further. The remaining heavy hadrons originating from the Big Bang should be present in
terrestrial matter. On the other hand, there are very strong upper limits on the abundances of nuclear isotopes
which contain such stable relics in the mass range from $1\,\mbox{GeV}$ to $10\,\mbox{TeV}$. Different experiments
set limits on their relative concentrations from $10^{-15}$ to $10^{-30}$ per nucleon \cite{42}.
At the same time, theoretical estimates show that if such remnant particles were to exist in nature today their
concentration should be much higher than $10^{-15}$ per nucleon \cite{43}.

Therefore the E$_6$CHM with stable
exotic colour triplets or stable exotic charged fermions is basically ruled out. In principle the set of exotic states
in the E$_6$CHM also includes neutral fermion states $\nu'$ $(\overline{\nu}')$ that transform non--trivially
under baryon triality. However, if these states are sufficiently light, i.e. they have masses in the few TeV range, to
play the role of dark matter such states would need to couple to the $Z$--boson and scatter on nuclei, resulting
in a spin--independent cross section which is a few orders of magnitude larger than the upper bound from
direct dark matter searches (for a recent analysis see \cite{Buckley:2013sca}).

In order to ensure that the composite Higgs model under consideration is phenomenologically viable,
we assume that the dynamics of the strongly interacting sector of the E$_6$CHM leads to the formation of the
$SU(6)$ singlet state $\overline{\eta}$, with spin $1/2$, which gains its mass through the mixing with the
elementary state $\eta$. In scenario A we allow for an interaction between $\overline{\eta}$,
vector $\Omega$ and ${\bf \overline{6}}_2$ of the type $\overline{\eta} \times \Omega \times {\bf \overline{6}}_2$.
We also assume that a similar interaction between $\overline{\eta}$, $\Omega$ and ${\bf \overline{6}}'_2$
is allowed in the case of scenario B. This implies that $\overline{\eta}$ carries baryon number $+1/3$ in
scenario A and $-1/3$ in scenario B. The breakdown of the EW symmetry gives rise to mixing between
$\overline{\eta}$ and $\overline{\nu}'$ as well as $\eta$ and $\nu'$, resulting in two mass eigenstates $\zeta_1$
and $\zeta_2$. When this mixing is rather small the lightest state, $\zeta_1$, can be predominantly
an $SU(2)_W$ singlet, so that its coupling to the $Z$--boson can be strongly suppressed. As a consequence
$\zeta_1$ can play the role of dark matter if this state is the lightest exotic state with non--zero $B_3$ charge.

When $\zeta_1$ is stable some part of the baryon asymmetry can be stored in the dark matter sector, because
$\zeta_1$ carries baryon number. Indeed, if $\zeta_1 \overline{\zeta}_1$ annihilation is efficient enough the
dark matter density in this model can be generated by the same mechanism that gives rise to the baryon
asymmetry of the Universe. In this case one can estimate the ratio of the baryon charges $B_{\zeta_1}$
and $B_n$ accumulated by $\zeta_1$ states and nucleons as
\be
\theta=\dfrac{B_{\zeta_1}}{B_n}\simeq \dfrac{1}{3} \left(\dfrac{\rho_{\zeta_1}}{\rho_n}\right) \left(\dfrac{m_n}{m_{\zeta_1}}\right)\,,
\label{ch32}
\ee
where $\rho_{\zeta_1}$ and $\rho_n$ are contributions of $\zeta_1$ states and nucleons to the total energy density,
while $m_{\zeta_1}$ and $m_n$ are the masses of the $\zeta_1$ states and nucleons, respectively. Taking into account that
$\rho_{\zeta_1}$ does not exceed the total dark matter density, i.e. $\rho_{\zeta_1}\lesssim 5 \rho_n$, the value of
$\theta$ can be larger than $0.1\%$ only when $m_{\zeta_1}\lesssim 2\,\mbox{TeV}$.

The presence of exotic states with TeV scale masses can lead to remarkable signatures. Assuming that $t'$, $b'_1$ and $e'_1$,
which stem from the same $SU(6)$ multiplet as the right--handed top quark, couple most strongly to the third generation fermions,
the vector--like exotic states $b'_1$ and $e'_1$ tend to decay into
\be
b'_1\to \overline{t}+\overline{b}+\zeta_1+X\,,\qquad\qquad e'_1\to  \overline{t}+b+\zeta_1+X\,.
\label{ch33}
\ee
The dominant decay channels of $b'_1$ and $e'_1$ are basically determined by the requirement of electromagnetic charge and
baryon number conservation. Since the exotic quark $t'$ can decay via $t'\to W^{*}+b^{'*}_1$, this exotic state results in a
similar final state to $b'_1$. In scenarios A and B the exotic quark $b'_2$ carries baryon number $+1/3$ and $-1/3$, respectively.
Thus in scenario A the decay channel
\be
b'_2\to Z+b
\label{ch34}
\ee
is allowed, whereas in scenario B this exotic state decays like $b'_1$ in scenario A. The exotic state $e'_2$ decays either via
$e'_2\to W+\overline{\zeta}_1$ (scenario A) or via $e'_2\to W+\zeta_1$ (scenario B).

If exotic quarks of the type described here do exist at sufficiently low scales, they can be accessed through direct pair
hadroproduction at the LHC. The corresponding production processes are generated via gluon--induced QCD interactions.
The exotic quarks $b'_1$ and $t'$ are doubly produced and decay into a pair of third generation quarks and $\zeta_1$,
resulting in the enhancement of the cross sections of
\be
pp\to t\overline{t}b\overline{b}+\Big/ \hspace{-0.35cm E_T}+X \qquad \mbox{and} \qquad
pp\to b\overline{b}b\overline{b}+\Big/ \hspace{-0.35cm E_T}+X\,.
\label{ch35}
\ee
The final states (\ref{ch35}) are similar to those associated with gluino pair production in the scenarios where the third generation
squarks are substantially lighter than the other sparticles, so that the gluino decays predominantly into a pair of third generation quarks
and a neutralino (for recent analysis see \cite{Barbieri:2009ev}). As compared with the exotic quarks, the direct production of $e'_1$, $e'_2$,
$\nu'$ and $\zeta_1$ is expected to be rather suppressed at the LHC. Nevertheless, it is worth noting that the pair production of
$e'_1 \overline{e}'_1$ can also lead to an enhancement of the cross sections for processes with the final states (\ref{ch35})
if $e'_1$ is sufficiently light.

Finally, let us consider the collider signatures associated with the scalar colour triplet $T$ that comes from the same pNGB $SU(5)$ multiplet,
$\tilde{H}$, as the composite Higgs doublet. In scenario A this scalar exotic state couples most strongly into $b'_2$ and $\zeta_1$.
Therefore, if $T$ is heavier than $b'_2$ it decays predominantly as:
\be
T\to b'_2 + \overline{\zeta}_1\,.
\label{ch36}
\ee
Otherwise it decays via
\be
T\to b + \overline{\zeta}_1 + X\,.
\label{ch37}
\ee
In scenario B the scalar colour triplet $T$ couples not only to $b'_2$ and $\zeta_1$ but also to $t$, $t'$, $b'_1$ and $e'_2$.
As a result, the following decay channels are allowed for this exotic state
\be
\ba{rcl}
T &\to & b'_2 (b'_1)+\zeta_1 \to  \overline{t}+\overline{b}+\zeta_1+\zeta_1+X\,,\\
T &\to & t' + e'_2 \to \overline{t}+\overline{b}+\zeta_1+\zeta_1+X\,,\\
T &\to & \overline{t} + \overline{b}'_2 \to \overline{t} + t + b +\overline{\zeta}_1 +X\,.
\ea
\label{ch38}
\ee
At the LHC, scalar colour triplets can be pair--produced if these exotic states are light enough. Then from Eq.~(\ref{ch38}) it follows
that the decays of $T\overline{T}$ may result in the enhancement of the cross sections for the processes (\ref{ch35}), with the four
third generation quarks in the final states. Besides, as one can also see from Eq.~(\ref{ch38}), in some cases the
$T\overline{T}$ production can lead to the enhancement of the cross sections that correspond to the processes with six
third generation quarks in the final states, i.e.
\be
pp \to  T\overline{T}\to t\overline{t}t\overline{t}b\overline{b}+\Big/ \hspace{-0.35cm E_T}+X\,,\qquad
pp \to  T\overline{T}\to b\overline{b}b\overline{b}b\overline{b}+\Big/ \hspace{-0.35cm E_T}+X\,.
\label{ch39}
\ee

\section{Conclusions}

In this paper we have studied a composite Higgs model which can arise naturally after the breakdown of
the $E_6$ gauge symmetry. Basically we focus on the GUT based on the $E_6\times G_0$ gauge group, which
is broken down to the $SU(3)_C\times SU(2)_W\times U(1)_Y \times G$ subgroup near some high energy scale $M_X$.
The low--energy limit of this GUT comprises strongly interacting and weakly--coupled sectors.
Gauge groups $G_0$ and $G$ are associated with the strongly coupled sector. Fields from this sector can be
charged under both $E_6$ and $G_0$ ($G$) gauge symmetries. The weakly--coupled sector involves elementary
states that participate in the $E_6$ interactions only. In this $E_6$ inspired composite Higgs model (E$_6$CHM)
all elementary quark and lepton fields can stem from the fundamental $27$-dimensional representation of $E_6$.

In order to avoid rapid proton decay and to guarantee the smallness of the Majorana masses of the left--handed
neutrino states, the Lagrangian of the strongly interacting sector of the E$_6$CHM has to be invariant under the
transformations of global $U(1)_B$ and $U(1)_L$ symmetries which ensure the conservation of the baryon and
lepton numbers to a very good approximation at low energies. Almost exact conservation of the $U(1)_B$ and $U(1)_L$
charges implies that elementary states with different baryon and/or lepton numbers must come from different
$27$--plets, while all other components of these multiplets gain masses of the order of $M_X$. Such a splitting of
the $E_6$ fundamental representations can occur within the six--dimensional orbifold SUSY GUT model presented in
this article. We consider the compactification of two extra dimensions on the orbifold
$T^2/(Z_2 \times Z^{I}_2 \times Z^{II}_2)$
that allows us to reduce the physical region to a pillow with four branes as corners. In this model the elementary quark
and lepton fields are components of different bulk $27$--plets, while all fields from the strongly coupled sector
are confined on the brane $O$, where $E_6$ symmetry is broken down to the $SU(6)\times SU(2)_N$ subgroup.
The $SU(6)$ group, that remains intact on the brane $O$, contains an $SU(3)_C\times SU(2)_W\times U(1)_Y$ subgroup.
We discuss the breakdown of the $E_6$ symmetry to the SM gauge group that results in the appropriate
splitting of the bulk $27$--plets. The $6D$ orbifold GUT models based on the $E_6$ gauge group do not have
an irreducible bulk anomaly, whereas brane anomalies get cancelled in the model under consideration.

In general the SM gauge couplings in the orbifold GUT models may not be identical near the scale $M_X$ where
the GUT gauge symmetry is broken. This is because sizable contributions to these couplings can come from
the branes where GUT symmetry is broken. Nevertheless, if the bulk contributions to the SM gauge couplings
dominate, approximate gauge coupling unification can take place. Since in the E$_6$CHM all states in
the strongly coupled sector fill complete $SU(6)$ representations, the convergence of the SM gauge couplings
is determined by the matter content of the elementary sector in the leading approximation. Then the approximate
unification of gauge couplings can be achieved if the right--handed top quark is entirely composite and the
weakly--coupled sector together with the SM fields (but without the right--handed top quark) contains a set of exotic states
so that its field content is given by Eq.~(\ref{ch2}). The presence of extra exotic states also ensures anomaly
cancellation in the elementary sector at low energies.

Since the strongly interacting sector is localised on the brane $O$, it can possess an $SU(6)\times SU(2)_N$ global symmetry
at high energies, even though local symmetry is broken down to the SM gauge group. In order to simplify our consideration
we assumed that $SU(2)_N$ symmetry is entirely broken. Thus the Lagrangian of the strongly coupled sector respects
$SU(6)\times U(1)_B\times U(1)_L$ global symmetry at high energies. The SM gauge interactions break $SU(6)$ global
symmetry. Nonetheless, if the gauge couplings of the strongly interacting sector are considerably larger than the SM gauge
couplings at any intermediate scale below $M_X$, then $SU(6)$ can be still an approximate global symmetry of the
composite sector at low energies.

We assumed that below scale $f\gg v$ the global $SU(6)$ symmetry is broken down to $SU(5)$, which includes
the $SU(3)_C\times SU(2)_W\times U(1)_Y$ subgroup. The $SU(6)/SU(5)$ coset space involves eleven pNGB states.
One of these pNGB states does not participate in the SM gauge interactions. Ten others form a fundamental representation
of $SU(5)$ with $5$ components. Two of these components are associated with the SM--like Higgs doublet $H$, while
three other components correspond to the $SU(3)_C$ triplet $T$. None of these pNGB states carry any baryon
and/or lepton numbers. The pNGB effective potential is induced by radiative corrections caused by the interactions between
elementary states and their composite partners that break $SU(6)$ symmetry. The structure of this scalar potential tends
to be such that it can give rise to the spontaneous breakdown of the EW symmetry, whereas $SU(3)_C$ colour is preserved.
Since the pNGB Higgs potential arises from loops, the effective quartic Higgs coupling tends to be sufficiently small that
it can lead to a $125\,\mbox{GeV}$ Higgs mass.

As in most composite Higgs models, the elementary quarks and leptons in the E$_6$CHM acquire their masses
through the mixing between these states and their composite partners. In particular, the corresponding masses can be generated
if all quark and lepton Yukawa couplings of the SM, which result in non--zero fermion masses, are allowed in the E$_6$CHM.
We argued that in the case of the quark sector this can happen in two different scenarios. Scenario A implies that the composite
partners of the left-handed quarks, the right-handed up-type and down-type quarks are components of ${\bf 20}$, ${\bf 15}$ and
${\bf \overline{15}}$ representations of $SU(6)$, respectively. In scenario B the composite partners of the right-handed up-type
quarks, left-handed quarks and right-handed down-type quarks belong to ${\bf 20}$, ${\bf 15}$ and ${\bf \overline{6}}$ representations
of the $SU(6)$ group. We also explored the generation of lepton masses. In both scenarios these masses can be induced if the
composite partners of the elementary left-handed leptons and right-handed charged leptons belong to ${\bf \overline{6}}$ and
${\bf 15}$ representations of $SU(6)$, respectively, whereas the composite partners of the right-handed neutrinos are
the $SU(6)$ singlet bound states. To ensure the smallness of the masses of the elementary left-handed neutrinos
we assumed that the elementary sector includes a set of heavy Majorana right-handed neutrino states with masses
somewhat below $M_X$, which do not participate in the $E_6$ gauge interactions but get mixed with the $SU(6)$ singlet
bound states that carry lepton number.

Because E$_6$CHM does not possesses any custodial symmetry, the electroweak precision observables are not protected
against the contributions of new composite states. As a result the stringent experimental constraint on the Peskin--Takeuchi $\hat{T}$
parameter pushes the $SU(6)$ symmetry breaking scale $f$ above $5-6\,\mbox{TeV}$. Besides, in the most general case
adequate suppression of the flavour--changing transitions in the E$_6$CHM requires $f\gtrsim 10\,\mbox{TeV}$.
The latter bound can be significantly relaxed if extra flavour symmetry is imposed. Nonetheless a significant fine--tuning,
$\sim 0.01\%$, is still needed to obtain the weak scale $v\ll f$. At first glance such model
might look a bit artificial. However the fact that no indication of new physics phenomena or any significant deviation from the SM
has been discovered at the LHC so far may suggest that the Higgs sector can be somewhat tuned. In other words, the scale of
new physics might be higher than previously thought.

The large value of the $SU(6)$ symmetry breaking scale also implies that the composite partners of the SM particles have
masses above $10\,\mbox{TeV}$, so that they are too heavy to be probed at the LHC. Moreover, since the deviations of the
couplings of the composite Higgs to the SM particles are determined by $v^2/f^2$, the modifications of the Higgs branching
fractions tend to be negligibly small in this model. So it seems rather problematic to test such small deviations at the LHC.
These small modifications of the Higgs branching ratios are probably even beyond the reach of a future $e^{+}e^{-}$ collider.
The couplings of the top quark to other SM particles are also expected to be extremely close to the ones predicted by the SM.

On the other hand, the spectrum of the E$_6$CHM contains one scalar colour triplet $T$ ($T^{\dagger}$) with electric
charge $-1/3$ ($+1/3$) and zero baryon number, as well as the set of vector--like fermions. All these states can have masses
in the few TeV range. The set of vector--like fermions, in particular, involves colour triplets $t' (\overline{t}')$ with electric charge
$+ 2/3 (-2/3)$, $b'_1$ and $b'_2$ ($\overline{b}'_1$ and $\overline{b}'_2$) with electric charge $- 1/3 (+1/3)$. The exotic quarks
$t'$ and $b'_1$ have baryon number $-1/3$. The colour triplet $\overline{b}'_2$ carries baryon numbers $-1/3$ and $+1/3$
in the scenarios A and B, respectively. To ensure the phenomenological viability of the model under consideration, the set
of exotic fermion states must include a Dirac fermion $\zeta_1$ ($\overline{\zeta}_1$), related to the exotic state
$\eta$ in Eq.~(\ref{ch2}), with baryon number $+ 1/3 (-1/3)$,
which is predominantly a SM singlet state. If such a fermion state were the lightest exotic particle, then it would tend to be stable
and could play the role of dark matter. The production cross sections of the colour triplet $T$ and exotic vector--like quarks  $t'$, $b'_1$
and $b'_2$ may not be negligibly small at the LHC provided these states are sufficiently light. Then the pair production of exotic
quarks may lead to the enhancement of the cross sections for $pp\to t\overline{t}b\overline{b}+\Big/ \hspace{-0.35cm E_T}+X$
and $pp\to b\overline{b}b\overline{b}+\Big/ \hspace{-0.35cm E_T}+X$\,. We also argued that in some cases the $T\overline{T}$ pair
production at the LHC can result in either similar final states with the four third generation quarks and missing energy or even
give rise to the enhancement of the cross sections that correspond to the processes with six third generation quarks and
missing energy in the final states, i.e. $pp \to  T\overline{T}\to t\overline{t}t\overline{t}b\overline{b}+\Big/ \hspace{-0.35cm E_T}+X$
and $pp \to  T\overline{T}\to b\overline{b}b\overline{b}b\overline{b}+\Big/ \hspace{-0.35cm E_T}+X$\,.

\section*{Acknowledgements}
R.N. is grateful to J.~Barnard, S.~Duplij, O.~Kancheli, D.~Kazakov, S.~F.~King, M.~M\"{u}hlleitner, M.~Sher, D.~G.~Sutherland, X.~Tata,
M.~Vysotsky for  fruitful discussions. This work was supported by the University of Adelaide and the Australian Research Council through
the ARC Center of Excellence in Particle Physics at the Terascale (CE 110001004) and through grant LF0 99 2247 (AWT).

\end{document}